\documentclass[%
  aps,
  superscriptaddress,
  notitlepage,
  onecolumn,
  bibnotes
]{revtex4-1}

\usepackage[utf8]{inputenc}
\usepackage[T1]{fontenc}

\usepackage{graphicx}
\usepackage{float}
\usepackage{adjustbox}
\usepackage{rotating}

\usepackage{booktabs}
\usepackage{tabularx}

\usepackage{amsmath,amsfonts}
\usepackage{siunitx}
\usepackage{gensymb}
\usepackage{textgreek}

\usepackage{soul}          
\usepackage[normalem]{ulem}
\usepackage{ragged2e}      
\usepackage{hyphenat}      
\usepackage{textcomp}      

\usepackage{enumitem}
\usepackage{changepage}
\usepackage{setspace}

\usepackage{lineno}

\usepackage{natbib}
\setcitestyle{numbers}  
\usepackage{natbib}
\usepackage[table,xcdraw]{xcolor}

\renewcommand{\cite}[1]{\textsuperscript{\citenum{#1}}}

\renewcommand{\doi}[1]{}
\renewcommand{\url}[1]{}

\makeatletter
\renewcommand\section{\@startsection {section}{1}{\z@}%
   {-2.0ex \@plus -1ex \@minus -.2ex}%
   {1.0ex \@plus.2ex}%
   {\color{cvblue} \large \bfseries}} 

\renewcommand\subsection{\@startsection {subsection}{2}{\z@}%
   {-1.5ex \@plus -1ex \@minus -.2ex}%
   {0.5ex \@plus .2ex}%
   {\color{cvblue} \normalfont\bfseries}} 

\makeatletter
  \renewcommand{\p@subsection}{}      
  \renewcommand{\p@subsubsection}{}   
\makeatother

\usepackage[colorlinks,linkcolor=cvblue,citecolor=cvblue,urlcolor=cvblue]{hyperref}

\usepackage{fancyhdr}
\definecolor{cvblue}{HTML}{0E5484}

\begin{document}
\setlength{\parskip}{0.55em}


\title{\large Revival of Layered Ferroelectrics in Thin Films}

\author{Elzbieta Gradauskaite}
\email{elzbieta.gradauskaite@cnrs-thales.fr}
\affiliation{Laboratoire Albert Fert, CNRS, Thales, Universit\'e Paris Saclay, 91767 Palaiseau, France \\ \normalfont{*}\href{mailto:elzbieta.gradauskaite@cnrs-thales.fr}{elzbieta.gradauskaite@cnrs-thales.fr}}

\date{\today}

\begin{abstract}
Layered perovskites are a versatile class of ferroelectrics with their structural anisotropy reflected in unusual electrostatics that give rise to exceptional ferroelectric properties. These materials fall into four main families: Aurivillius, Carpy–Galy, Ruddlesden–Popper, and Dion–Jacobson phases; each forming natural superlattices by interleaving perovskite slabs with spacer layers. For a long time, these materials were considered too structurally complex to prepare as high-quality thin films. However, recent breakthroughs in deposition and advanced characterization have made it possible to stabilize high-quality films with atomic precision, uncovering a wide range of unconventional ferroelectric functionalities. These include robust in-plane polarization without a critical thickness, the emergence of charged domain walls and non-trivial polar textures, resilience to doping with magnetic ions and charge carriers, and possibility to epitaxially integrate them into standard perovskite heterostructures. This review aims to unify current knowledge on the fabrication and characterization of layered ferroelectric thin films, and to present research findings across all four structural families, with the goal of highlighting their common features despite differences in crystal structure and polarization mechanisms. We also discuss promising research directions, including polar metallicity, (alter-)magnetoelectricity, exfoliation, and soft-chemistry-driven phase transformations, with the goal of consolidating the field and encouraging further exploration of these materials for both fundamental studies and applications.

\end{abstract}

\maketitle

  {\hypersetup{linkcolor=black}
   \tableofcontents
  }
\clearpage  


\section{Introduction}

Complex perovskite oxides are capable of stabilizing a broad range of physical phenomena, from magnetism\cite{bibes2007,bhattacharya2014, Trier2022} and ferroelectricity\cite{ahn2004a, muller2023} to multiferroicity\cite{Trassin2016a, fiebig2016b, Spaldin2019}, superconductivity\cite{bednorz1986,cheng2025}, and catalytic activity\cite{suntivich2011}. This functional diversity is underpinned by the chemical flexibility of the perovskite structure ($AB$O$_3$), which can accommodate a broad range of cations on both the $A$ and $B$ sites~\cite{Bartel2019}. The perovskite structure consists of a network of corner-sharing $B$O\textsubscript{6} octahedra, with the larger $A$-site cation at the corners of the (pseudo-)cubic unit cell. This simple yet versatile framework can easily distort, tilt, and adapt to non-stoichiometries, providing a powerful handle for tuning material properties. Beyond structural versatility, the richness of physical properties in perovskite oxides stems from the electronic configuration of the transition metal cations, particularly the spatially localized $d$- and $f$-orbitals. 

A breakthrough for oxide interface physics in the early 2000s was the realization that atomically sharp interfaces between oxide perovskites can give rise to emergent phenomena absent in the bulk. The most iconic example is the discovery of conductivity at the interface between two insulating oxides, LaAlO$_3$ and SrTiO$_3$\cite{ohtomo2004}. Other examples include charge transfer effects\cite{charge_transfer} and interfacial magnetoelectric coupling\cite{molegraaf2009, Vaz2012a, gradauskaite2024} mediated by strain or charge coupling. These discoveries triggered a surge of interest in artificially constructed superlattices, where the interfacial volume is maximized relative to the bulk, enabling interfacial effects to dominate the overall physical behavior.

Superlattice design has proven particularly fruitful in the field of ferroelectrics. By precisely tuning the periodicity (or ``wavelength'') of alternating ferroelectric and dielectric layers (e.g.\ PbTiO\textsubscript{3} and SrTiO\textsubscript{3}, respectively), researchers have achieved control over the ferroelectric state\cite{das2018}. The PbTiO\textsubscript{3} system, known for its single-domain out-of-plane polarization in epitaxial thin films\cite{Strkalj2020}, evolves with increasing periodicity through multidomain out-of-plane states\cite{Lichtensteiger2014a} to fully in-plane-polarized configurations\cite{Hong2021}. The delicate balance between elastic, electrostatic, and gradient energies can even stabilize hybrid improper ferroelectricity\cite{bousquet2008b} or non-trivial ferroelectric topologies such as flux-closure patterns\cite{Tang2015}, polar vortices\cite{Yadav2016,Hong2017}, three-dimensional domain crystals\cite{Stoica2019a}, and antipolar order\cite{yin2024}. All this is governed by the interplay between strain and electrostatic boundary conditions at each interface with dielectric layers. 

Nature provides an alternative route to such layered oxide architectures in the form of \textbf{``natural superlattices''}: crystal structures that exhibit periodic stacking driven purely by inherent electrostatic and bonding preferences, without the need for artificial layering or externally imposed strain. \textbf{Several families of such layered perovskite-based oxides exist, including Aurivillius (A), Carpy–Galy (CG), Ruddlesden–Popper (RP), and Dion–Jacobson (DJ) phases.} These compounds are composed of a variable number $n$ of perovskite-like planes interleaved with ionically distinct spacer layers along one crystallographic direction. This results in large unit cells with pronounced anisotropy, which often gives rise to emergent phenomena not seen in their perovskite counterparts. While most renowned for the superconductivity observed in the Ruddlesden--Popper phases\cite{maeno1994}, many of these layered compounds are also ferroelectric and exhibit a sizeable in-plane polarization\cite{benedek2015b}. 

This review focuses on layered perovskite-based ferroelectric oxides, with particular emphasis on recent experimental progress in thin-film synthesis and characterization. We begin by introducing the structural families and their associated polarization mechanisms. We then trace the evolution of thin-film growth strategies that enabled better crystalline quality and control over orientation. Characterization approaches specific to these materials are discussed, including defect analysis and their implications for ferroelectric functionality. This is followed by a survey of unconventional phenomena in layered ferroelectrics that emerge in the (epitaxial) thin-film limit. We conclude with a discussion on key opportunities, ranging from polar metallicity and multiferroicity to soft-chemistry methods for structural tuning, outlining exciting avenues for future research in layered perovskite-based ferroelectric oxides.

\section{Structural families and associated mechanisms for ferroelectricity}

There are four main families of layered perovskite-based ferroelectrics: Aurivillius, Carpy--Galy, Ruddlesden–Popper, and Dion–Jacobson. The Aurivillius ferroelectrics (also referred to as bismuth layered ferroelectrics (BLFs)) were first reported in 1949\cite{aurivilliusmain} and are by far the most studied, with the interest in these materials peaking in the 1990s due to their fatigue-free properties explored in SrBi$_2$Ta$_2$O$_9$ capacitors\cite{dearaujo1995a}. The Carpy–Galy family with La$_2$Ti$_2$O$_7$ as a prototypical example has been studied significantly less, even though ferroelectricity in these compounds was uncovered back in the 1970s\cite{carpy1974,galy1974}. One reason for that could be the lack of a clear unifying family name, as these materials were often referred to as perovskite-like layered structures (PLS) or simply as (110) layered ferroelectrics. It was not until 2019 that Núñez Valdez and Spaldin\cite{nunezvaldez2019b} insisted on the Carpy–Galy name, which will hopefully help to unify research efforts on this phase and render them more accessible. 

While Aurivillius and Carpy–Galy materials can be classified as proper ferroelectrics, ferroelectricity in Ruddlesden– Popper\cite{ruddlesden1957} and Dion–Jacobson\cite{dion1981, jacobson1985} phases is typically hybrid improper, i.e.\ two nonpolar distortions are the primary order parameters that induce ferroelectric polarization through trilinear coupling. Initially predicted from first-principles calculations in 2011 by Benedek and Fennie\cite{benedek2011}, hybrid improper ferroelectricity was first demonstrated in Ruddlesden–Popper compounds such as Ca$_3$Mn$_2$O$_7$ in 2015\cite{oh2015c}. Ferroelectricity in Dion--Jacobson compounds, such as RbNdNb\(_2\)O\(_7\), was theoretically predicted in 2006\cite{fennie2006}, experimentally confirmed in 2012\cite{li2012a}, and later explained by the unifying theory of hybrid improper ferroelectricity\cite{benedek2014b}. Hybrid–improper ferroelectricity has likewise been predicted for $A$-site–ordered double perovskites, (\textit{A},\textit{A}')\textit{B}\textsubscript{2}O\textsubscript{6}\cite{young2013, benedek2015b}. Because such cation-ordered structures are rarely found in nature, they are not discussed further in this review.

\begin{figure}[htb!]
  \centering 
  \begin{adjustbox}{width=0.75\textwidth, center}
    \includegraphics[width=0.75\textwidth,clip, trim=5 8 3 5]{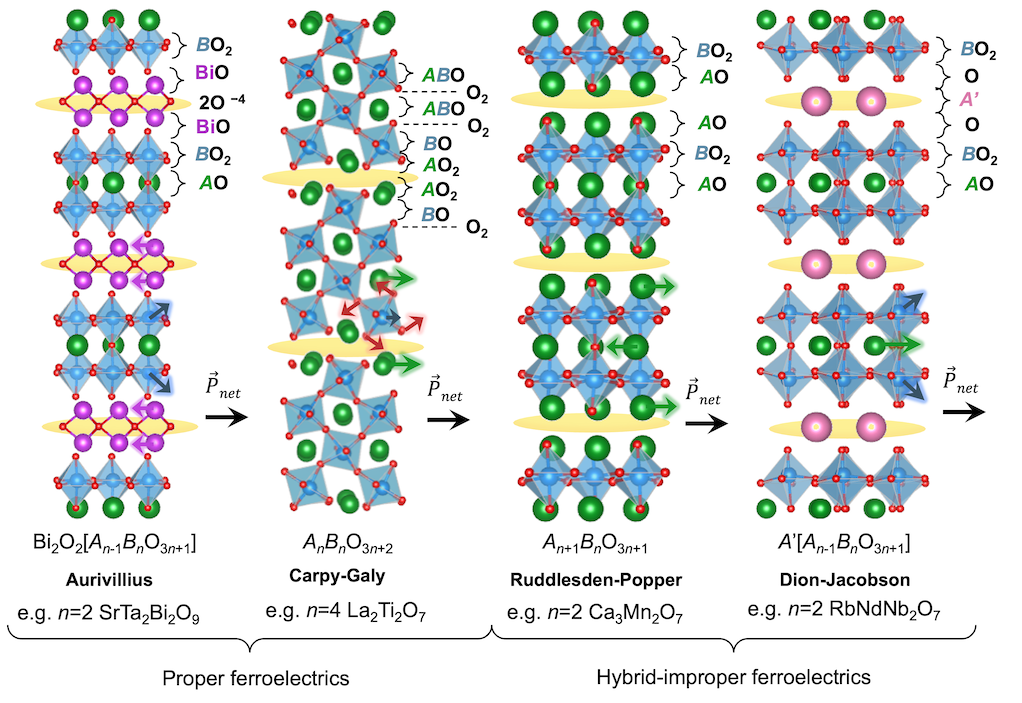}
  \end{adjustbox}
 \caption{\textbf{Structural families of layered ferroelectric perovskite-based oxides.} Schematic representations of four families: Aurivillius, Carpy--Galy, Ruddlesden–Popper, and Dion–Jacobson. Yellow highlights the ionic spacer layers, which concentrate charge and influence the polarization mechanisms. Colored arrows indicate typical ionic displacements driving ferroelectricity in each structure. Atomic plane labels on the right side denote the stacking sequence within each family.}
  \label{fig:families}
\end{figure}

We provide a brief overview of the structural differences between the four families, along with a short description of the polarization mechanisms characteristic of each. For a more detailed discussion of the crystal-chemical origins of polarization in these materials, we refer the reader to the seminal review by Benedek \textit{et al.}\cite{benedek2015b}.

\subsection{Aurivillius \texorpdfstring{$\mathrm{Bi}_2\mathrm{O}_2[\mathrm{A}_{n-1}\mathrm{B}_{n}\mathrm{O}_{3n+1}]$}{Aurivillius Bi2O2[An-1BnO3n+1]} compounds}\label{sec:A}
 
The unit cell of Aurivillius compounds\cite{aurivilliusmain,newnham1971a} is made up of alternating layers of a variable number \textit{n} of perovskite-octahedra planes interleaved between fluorite-like [Bi\textsubscript{2}O\textsubscript{2}]\textsuperscript{2+} layers along the \textit{c}-axis, see Figure \ref{fig:families}. Their high-symmetry phase is a tetragonal $I4/mmm$ structure that undergoes one or more\cite{hervoches2002a} transitions to a polar phase (e.g.\ orthorhombic $A2_{1}am$) upon cooling, characterized by large spontaneous in-plane polarization (30–50\,$\mu$C/cm$^2$)\cite{newnham1971a}.

The proper ferroelectricity in Aurivillius phases originates from in-plane distortions, where the [Bi\textsubscript{2}O\textsubscript{2}]\textsuperscript{2+} fluorite-like layers shear relative to the perovskite blocks\cite{newnham1971a,djani2012a,birenbaum2014a}. In addition to the ferroelectric soft mode, Aurivillius compounds exhibit additional non-polar distortions, such as in-plane tilts and out-of-plane rotations of the oxygen octahedra\cite{birenbaum2014a,benedek2015b}. Bi\textsuperscript{3+} cations exhibit lone-pair–driven off-center displacements along the crystallographic \textit{a}-axis (the in-plane polar axis), indicated by purple arrows in Figure~\ref{fig:families}. Within each perovskite slab, the oxygen octahedra tilt and rotate, and the $B$ cations shift off-center (blue arrows in Fig.~\ref{fig:families}). These $B$-site off-centerings adopt an antipolar-like order along the out-of-plane direction, while their in-plane components add up to produce net in-plane polarization antiparallel to the lateral Bi displacements in the fluorite-like layers. In compounds with even \textit{n}, the antipolar order leads to complete cancellation of the out-of-plane components, confining the polarization strictly to the in-plane \textit{a}-axis. In contrast, odd-\textit{n} compounds still exhibit dominant in-plane polarization but retain a small additional out-of-plane component due to incomplete cancellation across the odd number of layers\cite{Funakubo2008}.

Traditionally, the spacer layers in Aurivillius compounds are described as fluorite-like [Bi\textsubscript{2}O\textsubscript{2}]\textsuperscript{2+} layers, characterized by a more covalent bonding nature. However, by analogy with $A$O and $B$O\textsubscript{2} surface terminations in perovskite oxides\cite{Efe2021}, the two Bi layers within the fluorite-like plane can alternatively be interpreted as belonging to vertically distorted BiO ($A$O) planes of a perovskite structure, separated by an intermediate 2O\textsuperscript{4-} atomic plane. This negatively charged oxygen double layer (highlighted in yellow in Fig.~\ref{fig:families}) can significantly influence the electrostatic boundary conditions at interfaces between Aurivillius phases and other perovskite compounds\cite{Spaldin2021,gradauskaite2023,efe2025}.

\subsection{Carpy--Galy \texorpdfstring{$\mathrm{A}_{n}\mathrm{B}_{n}\mathrm{O}_{3n+2}$}{AnBnO3n+2} compounds}\label{sec:CG}

The layered Carpy–Galy ferroelectrics comprise a variable number \textit{n} of 110-oriented perovskite planes interleaved with additional oxygen layers\cite{isupov1999d, lichtenberg2001a, lichtenberg2008a} (Fig.\ \ref{fig:families}). The dominant ferroelectric instability in the Carpy--Galy phases originates from oxygen octahedral tilts/rotations\cite{lopez-perez2011a,nunezvaldez2019b}, which are commonly referred to as antiferrodistortive instabilities. In typical three-dimensional perovskite crystals, such antiferrodistortive modes create local electric dipoles that cancel each other out, resulting in zero net polarization. However, in  Carpy--Galy phases, the presence of oxygen spacer layers (highlighted in yellow) truncates the unit cell along the out-of-plane axis, creating an odd number of such electric dipoles between the spacers, which leads to a non-zero net in-plane polarization (typically up to 10 $\mu$C/cm$^2$ in bulk\cite{nanamatsuNewFerroelectricLa2Ti2o71974,nanamatsu1971} and shown to reach 18 $\mu$C/cm$^2$ in epitaxial thin films\cite{gradauskaite2025}). Because the polarization emerges from layering of the structure that creates this "dipole imbalance", the phenomenon is referred to as geometric or topological ferroelectricity\cite{lopez-perez2011a}. 

The proper ferroelectricity in these compounds is characterized by oxygen octahedral tilts/rotations, $A$-site lateral displacements, and slight $B$-cation offcentering, schematized by red, green, and blue arrows in Figure \ref{fig:families}, respectively. It is coupled to additional non-polar distortions, such as additional octahedral tilting or antipolar distortions\cite{gradauskaite2025}. Together, these distortions lower the symmetry to polar space groups such as \(P2_1\) or $Cmc2_1$ with a unique in-plane polarization axis. The multitude of polar and non-polar distortions present in Carpy–Galy compounds suggests numerous plausible phase-transition pathways between the high-temperature symmetric and ferroelectric space groups\cite{nunezvaldez2019b,gradauskaite2025}, which are challenging to disentangle, especially given that the $T_\mathrm{C}$ of these compounds is among the highest reported\cite{yan2009}. This also gives rise to flat phonon bands, which are linked to experimentally observed incommensurate intermediate phases\cite{capponia, daniels2002a, ohi1985c, howiesonIncommensurateCommensurateTransition2020}.

The highest polarization is typically observed for $n$=4 Carpy--Galy compounds with the empirical formula \textit{A}\textsubscript{2}\textit{B}\textsubscript{2}O\textsubscript{7} (nominally identical to that of pyrochlore \textit{A}\textsubscript{2}\textit{B}\textsubscript{2}O\textsubscript{7}\cite{hickox-young2022}, although their respective structural stabilities are governed by the size of the constituent cations\cite{shao2012a}). By analogy with $A$O and $B$O\textsubscript{2} surface terminations in perovskite oxides\cite{Efe2021}, the Carpy–Galy unit cell can be decomposed into $AB$O, O\textsubscript{2}, $B$O, and distorted $A$O\textsubscript{2} atomic planes, where the additional oxygen spacer (highlighted in yellow) can be represented as two adjacent $A$O\textsubscript{2} layers; see Figure~\ref{fig:families}.

\subsection{Ruddlesden--Popper \texorpdfstring{$\mathrm{A}_{n+1}\mathrm{B}_{n}\mathrm{O}_{3n+1}$}{An+1BnO3n+1} compounds}\label{sec:RP}

Ruddlesden–Popper phases comprise $n$ perovskite layers separated by a double-layer rock salt spacer (appears as two consecutive $A$O planes, see Figure \ref{fig:families}). In their high-symmetry parent structure, no unstable polar phonon exists; instead, two non-polar zone-boundary instabilities dominate: an in-phase octahedral rotation about the out-of-plane axis and an anti-phase tilt about an in-plane axis.\cite{benedek2011} These two non-polar modes enable the appearance of a polar mode as a result of a trilinear coupling\cite{benedek2011,benedek2012a,zhang2022a}. In this case, polarization is a secondary order parameter resulting from the product of the two primary rotations and tilts, in the so-called avalanche phase transitions, where all three modes condense together. The resulting polar displacement pattern in Ruddlesden--Popper ferroelectrics consists of lateral displacements of the $A$-site cations in each $A$O plane, which is represented by the green arrows in Figure \ref{fig:families}.  The direction of each displacement alternates from layer to layer.  For even $n$ the unit cell contains an odd number of $A$O planes within spacers, leading to a net in-plane polarization. Conversely, for odd $n$, there is a cancellation between dipoles yielding a non-polar ground state.   

Because polarization is not the primary order parameter, switching it requires reversing the octahedral rotations; consequently, the coercive fields are higher and the spontaneous polarization is generally lower than in proper ferroelectrics (up to 4 $\mu\mathrm{C\,cm^{-2}}$ \cite{zhang2022a}). There is a linear dependence between ferroelectric T\textsubscript{C} and Goldschmidt tolerance factor in these phases, confirming the purely geometric polarization origin\cite{yoshida2018c} and enabling rational design\cite{balachandran2014a, mulder2013b} polarization in $A$-site substituted compounds (\textit{A}, \textit{A'})\textit{B}\textsubscript{2}O\textsubscript{6}. On the one hand, to reduce the coercive field, the average tolerance factor of two perovskite layers (with $A$ and $A'$ cations) should be maximized\cite{mulder2013b}. On the other hand, the higher the difference in the tolerance factor between the two perovskite layers, the higher the polarization\cite{mulder2013b}. 

Recently, Markov \textit{et al.} identified a new structural family of anti–Ruddlesden–Popper phases, with reversed cation and anion locations, through the analysis of a high-throughput database of phonon band structures\cite{markov2021}.  In these compounds, the polarization is predicted to lie perpendicular to the layer stacking direction.

\subsection{Dion--Jacobson \texorpdfstring{$\mathrm{A}'[\mathrm{A}_{n-1}\mathrm{B}_{n}\mathrm{O}_{3n+1}]$}{A'[An-1BnO3n+1]} compounds}\label{sec:DJ}

The Dion–Jacobson phases consist of $n$ perovskite slabs separated by a large single $\mathrm{A}'$ cation layer instead of a double $A$O layer as in the Ruddlesden–Popper phases. $\mathrm{A}'$ is typically a univalent alkali cation. Despite this structural difference, the mechanism of ferroelectricity in Dion–Jacobson compounds is remarkably similar to that of Ruddlesden–Popper phases: the polarization arises from a trilinear coupling between two non-polar octahedral rotation/tilt modes\cite{benedek2014b}. Similarly to Ruddlesden--Popper ferroelectrics, those of the Dion–Jacobson family have moderate polarization (2-3 $\mu\mathrm{C\,cm^{-2}}$ \cite{asaki2020b}) and high switching barriers. Due to its geometric origins, polarization can be induced through cation exchange\cite{zhu2018}.

The hybrid–improper distortion drives lateral $A$-site shifts (green arrows in Fig.~\ref{fig:families}) and off-centre displacements of the $B$ cations within the $B$O$_{6}$ octahedra (blue arrows)\cite{li2012a}. First-principles study shows that in $n$=2 compounds, the $A$O layers supply a large share of the polarization, but the main contribution arises from the $B$O\textsubscript{2} layers (as there are more of them)\cite{benedek2014b}. 

\subsection{Common ferroelectric signatures: high T\textsubscript{C} and in-plane polarization}\label{sec:tc}

Ferroelectricity in these four phases is primarily stabilized by the layering and confinement effects of their large periodic unit cells rather than by electronic hybridization within the \textit{AB}O\textsubscript{3} units seen in conventional perovskite ferroelectrics like BaTiO\textsubscript{3}. Polarization stability determined on the larger scale means it often has superior stability and robustness compared to traditional ferroelectric perovskites. This is reflected in unusually high ferroelectric T\textsubscript{C} values: around 600--800\textdegree{}C in Aurivillius compounds\cite{fukunaga2016a, newnham1971a}, above 1000\textdegree{}C in Dion--Jacobson compounds\cite{li2012a, chen2015} and  as high as 800\textdegree{}C in Ruddlesden--Popper ferroelectrics\cite{kong2023a}, while with ca.\ 1500\textdegree{}C, Carpy--Galy compounds boast the likely highest T\textsubscript{C} of any known ferroelectric\cite{yan2009}.

All four families of layered perovskite ferroelectrics share a common structural motif: blocks of perovskite slabs separated by ionic spacer layers. During crystal or thin-film growth, these spacers usually align parallel to the substrate (or to the crystal facet with the largest surface area), since this orientation minimizes the surface energy\cite{treacy1990,nakajima2010}. Because the spacers carry a higher local ionic charge than the perovskite blocks, the structural anisotropy directly translates into anisotropy of the electrostatic boundary conditions. The charged layers preferentially screen the bound charges of the ferroelectric slabs, leading to cancellation of the out-of-plane polarization component (antipolar-like order). In contrast, the in-plane component remains uncompensated. Despite family-specific microscopic mechanisms, the net macroscopic polarization in all four layered systems is therefore typically uniaxial and lies in-plane, parallel to the spacer planes.

\subsection{Opportunities and challenges for integrating in-plane-polarized layered ferroelectrics into devices}

The in-plane polarization of naturally layered perovskite-based ferroelectrics presents both a fundamental feature and a practical challenge for device integration. In particular, it requires top electrodes with very small lateral separations: narrow gaps are necessary to generate the electric fields required to switch the polarization direction with relatively low voltages. With standard lithography techniques (e.g.\ projection lithography or electron-beam lithography), one is limited to gaps of ca.\ 1 \textmu m between interdigitated electrodes in order to fully transfer the features in the resist and achieve a complete lift-off\cite{alexanderliddle2002}. This is the reason why the fabrication of devices with in-plane ferroelectrics has long been deemed impractical. 

Early research attempted to circumvent this issue by forcing the layered unit cells to tilt with respect to the substrate surface normal, thereby reorienting the polarization to have a partial out-of-plane component\cite{lee2002, lee2003a, ullah2019b}. In parallel, first-principles studies searched for Ruddlesden–Popper phases with spontaneous polarization along the out-of-plane axis, identifying a few promising non-oxide candidates\cite{lu2023a}.

More recently, however, several developments have revived interest in exploiting the intrinsic in-plane polarization of these materials. Advances in extreme ultraviolet lithography (EUV) in the future will allow fabrication of nanoscale electrode features, enabling lateral gaps compatible with practical switching voltages and offering promising paths for future device architectures. In addition, all-optical methods for ferroelectric polarization switching and readout have been demonstrated\cite{lei2025, sarott2024}. It was also shown that in some ferroelectrics, nanoscale metallic regions can be induced by electron-beam-triggered phase transitions\cite{yao2020e,jang2025}, which could enable direct writing of metallic contacts in the same material matrix. These emerging strategies suggest that the integration of in-plane-polarized layered ferroelectrics into devices could one day become technologically feasible.

\section{Evolution of thin-film synthesis of layered ferroelectrics}

The layered ferroelectric families were all discovered as bulk single crystals, where functional ferroelectric domains and polarization were probed for the first time.  Device integration, however, demands thin films that are uniform over large areas, tolerant to heterogeneous stacking, and can be achieved through controlled synthesis (in terms of the correct phase, its orientation, and required thickness).  The fabrication of layered ferroelectrics as thin films started as early as the late 1960s, and since then the field has seen changes in synthesis techniques, choice of substrates, and characterization methods, that together have led to dramatic improvements of the thin film quality and associated functional properties.

\begin{figure}[b!]
  \centering 
  \begin{adjustbox}{width=0.92\textwidth, center}
    \includegraphics[width=0.92\textwidth,clip, trim=5 5 3 5]{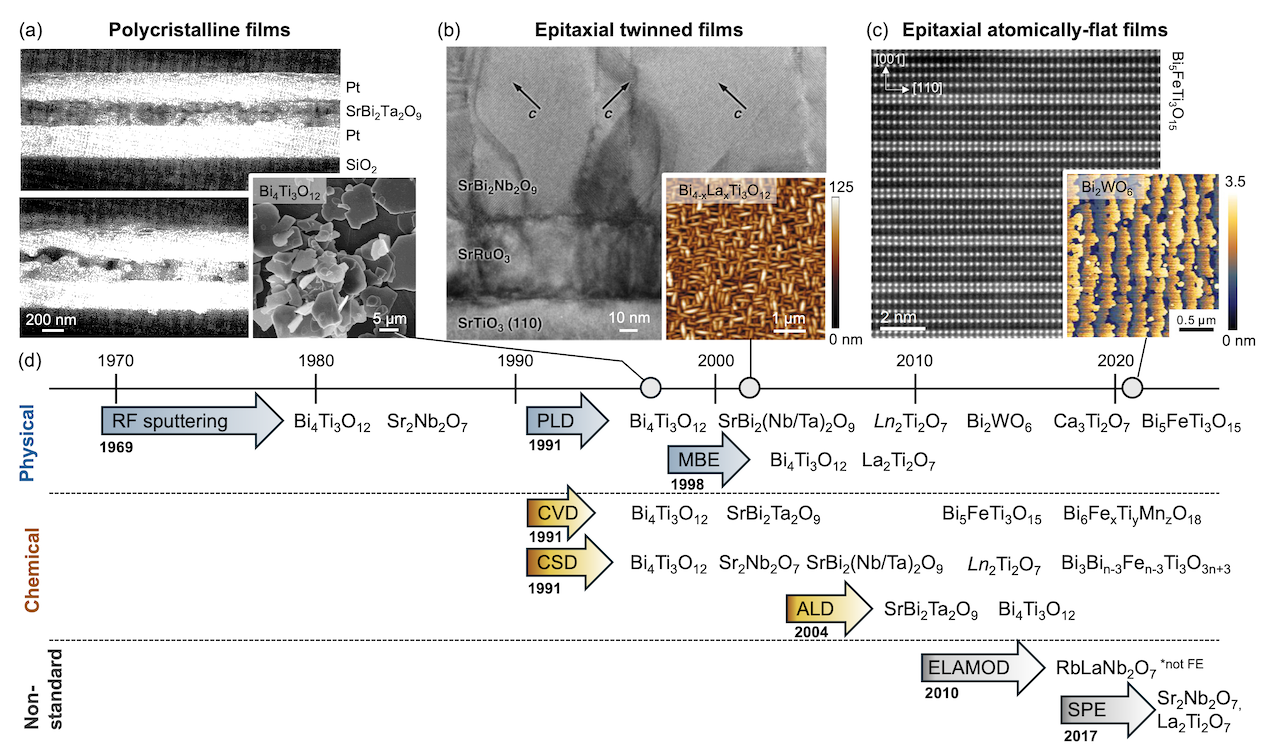}
  \end{adjustbox}
  \caption{\textbf{Historical evolution of deposition techniques and improvement in layered ferroelectric thin-film quality.} 
  Representative layered ferroelectric oxides synthesized as thin films are shown across three stages of development: 
  (a) Polycrystalline films obtained by CSD methods, as seen in cross-sectional scanning electron microscopy (SEM) micrographs of a typical SrBi\textsubscript{2}Ta\textsubscript{2}O\textsubscript{9} capacitor stack before and after high-temperature annealing~\cite{noguchi1996}. Reproduced from~\citenum{noguchi1996} with permission from IOP Publishing. Scanning electron microscopy (SEM) image (inset) revealing large Bi\textsubscript{4}Ti\textsubscript{3}O\textsubscript{12} crystallites prepared using molten-salt synthesis~\cite{horn1999}. Reproduced with permission from~\citenum{horn1999}. Copyright 1999 Wiley‐VCH GmbH.  
  (b) Epitaxial twinned films grown by PLD, showing tilted crystallographic domains in cross-sectional transmission electron microscopy (TEM) of SrBi\textsubscript{2}Nb\textsubscript{2}O\textsubscript{9}~\cite{lettieri}. Reproduced with permission from~\citenum{lettieri}. Copyright 2000 AIP Publishing. La-substituted SrBi\textsubscript{2}Ta\textsubscript{2}O\textsubscript{9} topography with anisotropic grain structures~\cite{lee2002}. Adapted with permission from~\citenum{lee2002}. Copyright 2002 The American Association for the Advancement of Science.  
  (c) Atomically resolved layering in Aurivillius Bi\textsubscript{5}FeTi\textsubscript{3}O\textsubscript{15} $n$=4 films~\cite{gradauskaite2021b}. Adapted with permission from~\citenum{gradauskaite2021b}. Copyright 2021 American Chemical Society. Atomically flat terraces in Bi\textsubscript{2}WO\textsubscript{6} films~\cite{das2021b}, demonstrating quality comparable to epitaxial perovskite oxide films. Adapted with permission from~\citenum{das2021b}. Copyright 2021 American Chemical Society.  
  (d) The timeline outlines the emergence of key thin-film-deposition methods grouped into physical, chemical, and non-standard approaches and lists representative compositions successfully stabilized by each; see Table~\ref{tab:fel_films_portrait} for the study references.}
   \label{fig:synthesis}
\end{figure}

\subsection{Technique evolution: from ceramic films to modern epitaxy}
The first report of layered ferroelectrics synthesized as thin films dates back to 1969, when Takei \emph{et al.}\cite{takei1969} used RF-sputtering to produce Aurivillius Bi\textsubscript{4}Ti\textsubscript{3}O\textsubscript{12} thin films on Pt and MgO. In the 1970s, sputtering was also utilized to obtain Carpy--Galy Sr\textsubscript{2}Nb\textsubscript{2}O\textsubscript{7} films\cite{ishitani1976}. It was not until 1990s that three independent advances arrived almost simultaneously.  Ramesh \emph{et al.}\ first used pulsed-laser deposition (PLD) to grow Aurivillius Bi\textsubscript{4}Ti\textsubscript{3}O\textsubscript{12} directly on epitaxial cuprate electrodes, demonstrating the feasibility of epitaxy on silicon wafers via an oxide buffer\cite{ramesh1991a,ramesh1991b}.  In the same year, Joshi \emph{et al.}\ introduced a sol-gel process, one of the chemical-solution deposition (CSD) approaches, to produce Bi\textsubscript{4}Ti\textsubscript{3}O\textsubscript{12}\cite{joshi1991}.  Finally, Wills \emph{et al.} used organometallic chemical vapor deposition (CVD) to deliver Aurivillius films\cite{wills1991}. At that time, films were typically a few hundred nanometers thick and consisted of large, misoriented grains. Post‐annealing treatments were therefore required, but these often introduced additional surface roughness\cite{noguchi1996,horn1999} (Fig.~\ref{fig:synthesis}a). The fatigue-free behavior in Aurivillius SrBi\textsubscript{2}Ta\textsubscript{2}O\textsubscript{9} capacitors demonstrated in 1995\cite{dearaujo1995a} fueled even more research efforts in the field. Molecular-beam epitaxy (MBE) was first used in 1998, improving upon phase purity in Aurivillius films\cite{theis1998}. Atomic layer deposition (ALD) soon followed: plasma-enhanced ALD of SrBi\textsubscript{2}Ta\textsubscript{2}O\textsubscript{9}  enabled the thin-film deposition with higher conformality and at lower temperatures\cite{shin2004, vehkamaki2006}. Figure \ref{fig:synthesis}d shows the chronological development of deposition methods used to prepare layered ferroelectric thin films.

A more detailed overview of reported layered ferroelectric perovskite-based oxide thin films is provided in Table~\ref{tab:fel_films_portrait}, summarizing key materials across the Aurivillius, Carpy–Galy, Ruddlesden–Popper, and Dion–Jacobson families, which have been synthesized as thin films. Over the years, PLD has been the preferred physical vapour deposition method, with most reported layered ferroelectrics grown using it. Among chemical deposition methods, CVD and sol-gel have become standard synthesis approaches for these materials.

{\setlength{\extrarowheight}{4.5pt}
\begin{table}[!htbp]
  \centering
  \small
  \caption{\textbf{Overview of reported layered ferroelectric perovskite‐based oxide thin films}: crystal family (A = Aurivillius; CG = Carpy–Galy; RP = Ruddlesden–Popper; DJ = Dion–Jacobson), number $n$ of perovskite layers, substrates, growth techniques, film quality, remanent polarization $P$, coercive field $E_\mathrm{c}$, domain imaging method, and associated references. Substrates: SrTiO$_3$ (STO), LaAlO$_3$ (LAO), DyScO$_3$ (DSO), NdGaO$_3$ (NGO), (LaAlO$_3$)$_{0.3}$(Sr$_2$AlTaO$_6$)$_{0.7}$ (LSAT).}
  \label{tab:fel_films_portrait}
  \resizebox{\textwidth}{!}{%
    \begin{tabular}{llllllllll}
      \hline
      Material & Class & $n$ & Substrate & Technique & Film quality & $P$ (\si{\micro C/cm^2}) & $E_\mathrm{c}$ (kV/cm) & Domains & Ref. \\
      \hline
      \parbox[t]{3cm}{SrBi$_2$Nb$_2$O$_9$} 
        & A & 2 
        & \parbox[t]{2.5cm}{STO (110), STO (001)} 
        & \parbox[t]{2cm}{PLD, sol–gel} 
        & \parbox[t]{2.2cm}{epitaxial, twinned} 
        & 22.8 & 200 
        & -- 
        & \parbox[t]{1.5cm}{\citenum{lettieri,zurbuchen2003a,boulle2005a}} \\[3pt]

      \parbox[t]{3cm}{SrBi$_2$Ta$_2$O$_9$} 
        & A & 2 
        & \parbox[t]{2.5cm}{Pt/TiO$_x$/SiO$_2$/Si, graphite} 
        & \parbox[t]{2cm}{CSD, MOCVD, ALD, PLD} 
        & \parbox[t]{2.2cm}{polycrystalline} 
        & -- & -- 
        & -- 
        & \parbox[t]{1.5cm}{\citenum{amanuma1995,dearaujo1995a,noguchi1996,burgess2000,shin2004, golosov2016a, lee2024}} \\[3pt]

      \parbox[t]{3cm}{SrBi$_2$Ta$_2$O$_9$} 
        & A & 2 
        & \parbox[t]{2.5cm}{LAO (001), STO (001)} 
        & \parbox[t]{2cm}{PLD} 
        & \parbox[t]{2.2cm}{epitaxial, twinned} 
        & -- & -- 
        & -- 
        & \parbox[t]{1.5cm}{\citenum{zurbuchen2003a}} \\[3pt]
        
    \parbox[t]{3cm}{Sr$_5$Bi$_4$Ti$_8$O$_{27}$} 
        & A & 8
        & \parbox[t]{2.5cm}{STO (001)} 
        & \parbox[t]{2cm}{PLD} 
        & \parbox[t]{2.2cm}{epitaxial} 
        & -- & -- 
        & -- 
        & \parbox[t]{1.5cm}{\citenum{zurbuchen2012}} \\[3pt]

      \parbox[t]{3cm}{Bi$_4$Ti$_3$O$_{12}$} 
        & A & 3 
        & \parbox[t]{2.5cm}{quartz; Pt/SiO$_2$/Si; sapphire} 
        & \parbox[t]{2cm}{sol–gel, MOCVD, ALD} 
        & \parbox[t]{2.2cm}{polycrystalline} 
        & 0.5\,(ALD) & -- 
        & -- 
        & \parbox[t]{1.5cm}{\citenum{joshi1991,wills1991,li1996,watanabe2000,vehkamaki2006}} \\[3pt]

      \parbox[t]{3cm}{Bi$_4$Ti$_3$O$_{12}$} 
        & A & 3 
        & \parbox[t]{2.5cm}{Pt; STO (001); LAO (001); MgAl$_2$O$_4$} 
        & \parbox[t]{2cm}{rf–sputtering, PLD, MBE, CVD} 
        & \parbox[t]{2.2cm}{epitaxial, twinned} 
        & -- & -- 
        & PFM 
        & \parbox[t]{1.5cm}{\citenum{takei1969,ramesh1991a,ramesh1991b,theis1998,deepak2013c}} \\[3pt]

      \parbox[t]{3cm}{La-doped\\ Bi$_4$Ti$_3$O$_{12}$} 
        & A & 3 
        & \parbox[t]{2.5cm}{SrRuO$_3$/YSZ/Si(100)} 
        & \parbox[t]{2cm}{PLD} 
        & \parbox[t]{2.2cm}{epitaxial, twinned, minor misorientation} 
        & 32 & 265 
        & -- 
        & \parbox[t]{1.5cm}{\citenum{lee2002,lee2003a}} \\[3pt]

      \parbox[t]{3cm}{Bi$_2$WO$_6$} 
        & A & 1 
        & \parbox[t]{2.5cm}{STO (100), LAO (100)} 
        & \parbox[t]{2cm}{PLD} 
        & \parbox[t]{2.2cm}{epitaxial, twinned} 
        & -- & -- 
        & PFM, STM 
        & \parbox[t]{1.5cm}{\citenum{wang2016c,jeong2021b, das2021b, er2019a, kwon2023, ullah2019b}} \\[3pt]

      \parbox[t]{3cm}{Bi$_5$FeTi$_3$O$_{15}$} 
        & A & 4 
        & \parbox[t]{2.5cm}{Pt/Si; Si(100); NGO (001); STO (100)} 
        & \parbox[t]{2cm}{CSD, PLD, AVD} 
        & \parbox[t]{2.2cm}{polycrystalline, oriented} 
        & 20.8, 16.5 & 250 
        & PFM, STM 
        & \parbox[t]{1.5cm}{\citenum{song2018a,zhang2012a,gradauskaite2020,gradauskaite2021b,gradauskaite2023,deepak2015a,efe2025,gradauskaite2022}} \\[3pt]

      \parbox[t]{3cm}{Bi$_6$Fe$_2$Ti$_3$O$_{18}$} 
        & A & 5 
        & \parbox[t]{2.5cm}{Pt/Si; sapphire} 
        & \parbox[t]{2cm}{CSD, AVD} 
        & \parbox[t]{2.2cm}{polycrystalline} 
        & 24 & 300 
        & -- 
        & \parbox[t]{1.5cm}{\citenum{song2018a,faraz2015b}} \\[3pt]

      \parbox[t]{3cm}{Bi$_7$Fe$_3$Ti$_3$O$_{21}$} 
        & A & 6 
        & \parbox[t]{2.5cm}{Pt/Si; NGO (001)} 
        & \parbox[t]{2cm}{CSD, PLD} 
        & \parbox[t]{2.2cm}{polycrystalline} 
        & 22.4 & 300 
        & PFM, STM 
        & \parbox[t]{1.5cm}{\citenum{song2018a,gradauskaite2021b}} \\[3pt]

      \parbox[t]{3cm}{Bi$_9$Fe$_5$Ti$_3$O$_{27}$} 
        & A & 8 
        & \parbox[t]{2.5cm}{NGO (001)} 
        & \parbox[t]{2cm}{PLD} 
        & \parbox[t]{2.2cm}{polycrystalline} 
        & -- & -- 
        & PFM, STM 
        & \parbox[t]{1.5cm}{\citenum{song2018a,gradauskaite2021b}} \\[3pt]

      \parbox[t]{3cm}{Bi$_6$Ti$_x$Fe$_y$Mn$_z$O$_{18}$} 
        & A & 5 
        & \parbox[t]{2.5cm}{NGO (001), LSAT (100), STO (100)} 
        & \parbox[t]{2cm}{DLI-CVD} 
        & \parbox[t]{2.2cm}{epitaxial / polycrystalline} 
        & -- & -- 
        & PFM 
        & \parbox[t]{1.5cm}{\citenum{keeney2020d, keeney2022, keeney2017a, moore2022b}} \\[3pt]

      \parbox[t]{3cm}{BaBi$_4$Ti$_4$O$_{15}$} 
        & A & 4 
        & \parbox[t]{2.5cm}{Nb:STO (001)} 
        & \parbox[t]{2cm}{PLD} 
        & \parbox[t]{2.2cm}{epitaxial, twinned} 
        & -- & -- 
        & PFM 
        & \parbox[t]{1.5cm}{\citenum{ahn2021a}} \\[3pt]

      \parbox[t]{3cm}{La$_2$Ti$_2$O$_7$} 
        & CG & 4 
        & \parbox[t]{2.5cm}{STO (110), STO (100), LSAT (110), DSO (100)} 
        & \parbox[t]{2cm}{MBE, PLD, sol–gel} 
        & \parbox[t]{2.2cm}{epitaxial single-crystal / twinned} 
        & 18.3 & 16.9 
        & PFM, STEM 
        & \parbox[t]{1.5cm}{\citenum{seo1998a,seo2001c,fompeyrinea,havelia2008,shao2011a,kaspar2018b,qiao2022b,gradauskaite2025,bayart2019c,havelia2009c,shao2012a}} \\[3pt]

      \parbox[t]{3cm}{Sr$_2$Nb$_2$O$_7$} 
        & CG & 4 
        & \parbox[t]{2.5cm}{Sr$_2$Ta$_2$O$_7$ (single crystal)} 
        & \parbox[t]{2cm}{rf–sputtering} 
        & \parbox[t]{2.2cm}{oriented} 
        & -- & -- 
        & -- 
        & \parbox[t]{1.5cm}{\citenum{ishitani1976a}} \\[3pt]

      \parbox[t]{3cm}{Sr$_2$Nb$_2$O$_7$} 
        & CG & 4 
        & \parbox[t]{2.5cm}{Si(100), Pt-coated Si(100)} 
        & \parbox[t]{2cm}{sol–gel} 
        & \parbox[t]{2.2cm}{polycrystalline} 
        & -- & -- 
        & -- 
        & \parbox[t]{1.5cm}{\citenum{prasadarao}} \\[3pt]

      \parbox[t]{3cm}{Sr$_2$Nb$_2$O$_7$} 
        & CG & 4 
        & \parbox[t]{2.5cm}{STO (110), LAO (110)} 
        & \parbox[t]{2cm}{PLD, solid-phase epitaxy} 
        & \parbox[t]{2.2cm}{epitaxial} 
        & -- & -- 
        & -- 
        & \parbox[t]{1.5cm}{\citenum{balasubramaniam2008,nezu2017,yao2020e}} \\[3pt]

      \parbox[t]{3cm}{Nd$_2$Ti$_2$O$_7$} 
        & CG & 4 
        & \parbox[t]{2.5cm}{STO (110)} 
        & \parbox[t]{2cm}{PLD, sol–gel} 
        & \parbox[t]{2.2cm}{epitaxial} 
        & -- & -- 
        & -- 
        & \parbox[t]{1.5cm}{\citenum{bayart2013a,bayart2019c,shao2012a,carlier2018a}} \\[3pt]

      \parbox[t]{3cm}{Pr$_2$Ti$_2$O$_7$} 
        & CG & 4 
        & \parbox[t]{2.5cm}{STO (110)} 
        & \parbox[t]{2cm}{PLD, sol–gel} 
        & \parbox[t]{2.2cm}{epitaxial} 
        & -- & -- 
        & -- 
        & \parbox[t]{1.5cm}{\citenum{bayart2019c,shao2012a}} \\[3pt]

      \parbox[t]{3cm}{Ce$_2$Ti$_2$O$_7$} 
        & CG & 4 
        & \parbox[t]{2.5cm}{STO (110)} 
        & \parbox[t]{2cm}{PLD, sol–gel} 
        & \parbox[t]{2.2cm}{epitaxial} 
        & -- & -- 
        & -- 
        & \parbox[t]{1.5cm}{\citenum{bayart2019c,shao2012a}} \\[3pt]

      \parbox[t]{3cm}{Sm$_2$Ti$_2$O$_7$} 
        & CG & 4 
        & \parbox[t]{2.5cm}{STO (110)} 
        & \parbox[t]{2cm}{PLD, sol–gel} 
        & \parbox[t]{2.2cm}{epitaxial} 
        & -- & -- 
        & -- 
        & \parbox[t]{1.5cm}{\citenum{bayart2019c,shao2012a}} \\[3pt]

      \parbox[t]{3cm}{Ca$_3$Ti$_2$O$_7$} 
        & RP & 2 
        & \parbox[t]{2.5cm}{STO (110)} 
        & \parbox[t]{2cm}{PLD} 
        & \parbox[t]{2.2cm}{epitaxial, twinned} 
        & 8\,(at 2K) & 5\,(at 2K) 
        & -- 
        & \parbox[t]{1.5cm}{\citenum{li2017b}} \\[3pt]

      \parbox[t]{3cm}{Ca$_3$Mn$_2$O$_7$} 
        & RP & 2 
        & \parbox[t]{2.5cm}{STO (100)} 
        & \parbox[t]{2cm}{PLD} 
        & \parbox[t]{2.2cm}{polycrystalline} 
        & -- & -- 
        & -- 
        & \parbox[t]{1.5cm}{\citenum{silva2023}} \\[3pt]

      \parbox[t]{3cm}{CsBiNb$_2$O$_7$} 
        & DJ & 2 
        & \parbox[t]{2.5cm}{LAO (001); free‐standing} 
        & \parbox[t]{2cm}{molten‐salt‐assisted synthesis} 
        & \parbox[t]{2.2cm}{epitaxial} 
        & -- & -- 
        & STEM 
        & \parbox[t]{1.5cm}{\citenum{cahill2010,guo2021b}} \\[3pt]
      \hline
    \end{tabular}%
  }
\end{table}
} 

\subsection{Improvement in thin-film textures through epitaxy}

The 2000s saw improvements in thin-film textures as researchers transitioned from growing layered ferroelectrics on Pt/Si/glass wafers to using single-crystalline perovskite oxide substrates. This led to epitaxially oriented films, which enabled controlling the majority-grain orientation and therefore increased net switchable polarization in the films\cite{lettieri,lee2002}. However, since anisotropic unit cells of layered ferroelectrics cannot be fully epitaxially matched by cubic perovskite substrates (Sec. \ref{sec:XRD}), the films still suffered from crystallographic twins and rather grainy structures, see Figure \ref{fig:synthesis}b. True single-crystallinity in epitaxial films emerged only when both lattice- and symmetry-matched substrates became available and were chosen for synthesis. For instance, Aurivillius Bi\textsubscript{\textit{n}+1}Fe\textsubscript{\textit{n}-3}Ti\textsubscript{3}O\textsubscript{3\textit{n}+3} films were grown by PLD on NdGaO\textsubscript{3} substrates cut along the orthorhombic (001) plane (the more common NdGaO\textsubscript{3} cut is pseudocubic (110)). This approach yielded twin-free, single-crystalline films with uniaxial in-plane polarization and atomically flat surfaces\cite{gradauskaite2020,gradauskaite2021b} (Fig.\ \ref{fig:synthesis}c). Aside from improving thin-film quality and reducing crystallographic twinning, epitaxial strain has also been predicted to serve as an additional tuning knob for polarization in layered ferroelectrics. Theoretically, it has been shown to drive phase transitions between polar, non-polar, and antiferroelectric states\cite{lu2016a, shah2010a, zhang2017d, cui2024a}.

\subsection{Non-conventional synthesis methods driven by structural anisotropy}

While the anisotropic stacking of layered perovskites is often viewed as a crystal-growth challenge, that same anisotropy can be leveraged for unconventional growth strategies. One such approach is solid-phase epitaxy (SPE), where an amorphous or metastable precursor film is deposited at low temperature and subsequently annealed. This method has proven effective for stabilizing kinetically demanding Carpy--Galy layered phases (e.g.\ Sr\textsubscript{2}Nb\textsubscript{2}O\textsubscript{7} and La\textsubscript{2}Ti\textsubscript{2}O\textsubscript{7})\cite{havelia2009c, bayart2013a}. When annealed in an oxygen atmosphere, these materials crystallize with their spacer layers preferentially aligned parallel to the substrate\cite{nezu2017, kaspar2018b, gradauskaite2025}. This behavior contrasts with isotropic cubic perovskites, where recrystallization typically leads to randomly oriented grains and twinning. In layered ferroelectrics, by contrast, the inherent structural anisotropy acts as a self-templating guide, enabling the recovery of epitaxial alignment, hence the term “solid-phase epitaxy.” Nevertheless, multi-site nucleation during annealing can still result in stacking faults, grain boundaries, and surface roughness.

This limitation can be addressed by an alternative strategy known as excimer-laser-assisted metal-organic deposition (ELAMOD), developed by Nakajima and co-workers\cite{nakajima2011, nakajima2014b, nakajima2014c}. ELAMOD similarly capitalizes on the structural anisotropy of layered materials, but initiates crystallization through localized energy input. A pulsed excimer laser selectively heats only the film surface, triggering crystallization exclusively from a top surface. The resulting growth front then gradually propagates through the film thickness, maintaining uniform orientation and significantly reducing the likelihood of stacking defects. Although demonstrated on non-ferroelectric Dion–Jacobson compounds such as RbLaNb\textsubscript{2}O\textsubscript{7} and RbCa\textsubscript{2}Nb\textsubscript{3}O\textsubscript{10}, the method shows how anisotropic structure can be exploited to achieve directional, self-aligned film growth even on non-lattice matching substrates like glass\cite{nakajima2014b}.

\section{Structural \& functional characterization of layered\\ perovskite-based ferroelectrics }

The structural and functional characterization of layered perovskite-based ferroelectrics builds on techniques established for conventional perovskites, but their large unit cells introduce additional challenges. For instance, reflection high-energy electron diffraction (RHEED) can track growth dynamics, yet the large periodicities of layered oxides often lead to multiple oscillations per unit cell that complicate interpretation (Sec.\ \ref{sec:in_situ_growth}). In X-ray diffraction (XRD), the large unit cells produce many additional reflections compared to simple perovskites, as well as artefacts related to stacking defects, increasing the complexity of the data (Sec.\ \ref{sec:XRD}). In scanning transmission electron microscopy (STEM), mapping polarization becomes more involved because several distortion modes may contribute simultaneously, rather than just the simple $B$-site off-centering familiar from perovskites (Sec.\ \ref{sec:STEM}). Finally, in scanning probe microscopy, the strong structural anisotropy means that crystal orientation critically affects the measurement, with some geometries providing better access to polarization (Sec.\ \ref{sec:pfm}). In this part, we review how these techniques are applied to layered perovskite-based ferroelectrics, with a focus on the specific considerations and adaptations they require.

\subsection{\textit{In-situ} diagnostics}\label{sec:in_situ_growth}

The superlattice structure of layered perovskite-based ferroelectrics forms spontaneously due to unit-cell electrostatics, without the need for specific epitaxial strain or manual layering. This represents a clear advantage over artificial superlattices such as (PbTiO$_3$/SrTiO$_3$)$_n$ \cite{Tang2015,Yadav2016,Hong2017}. As a result, the deposition of layered ferroelectric films can be achieved using a single precursor with the correct stoichiometry. Provided that the appropriate growth parameters are chosen to access the thermodynamic stability window of the desired phase, one can expect to stabilize the correct layered structure. For example, PLD growth uses a single target with the desired stoichiometry, making the synthesis efficient and straightforward. However, techniques such as MBE and CVD might still rely on an alternating sequence of precursors. To achieve Bi$_4$Ti$_3$O$_{12}$ by MBE, Bi and Ti fluxes are introduced separately during growth\cite{theis1998}, whereas CVD growth involves timed injections of Bi(thd)$_3$ (thd = 2,2,6,6-tetramethyl-3,5-heptanedionate) and Ti(O-iPr)$_2$(thd)$_2$ (where O-iPr = iso-propoxide) solutions\cite{deepak2013c}. For some growth methods that allow \textit{in-situ} diagnostics, the first characterization of layered ferroelectric thin films can already be carried out during deposition.

\begin{figure}[htb!]
  \centering 
  \begin{adjustbox}{width=0.96\textwidth, center}
    \includegraphics[width=0.96\textwidth,clip, trim=5 5 3 5]{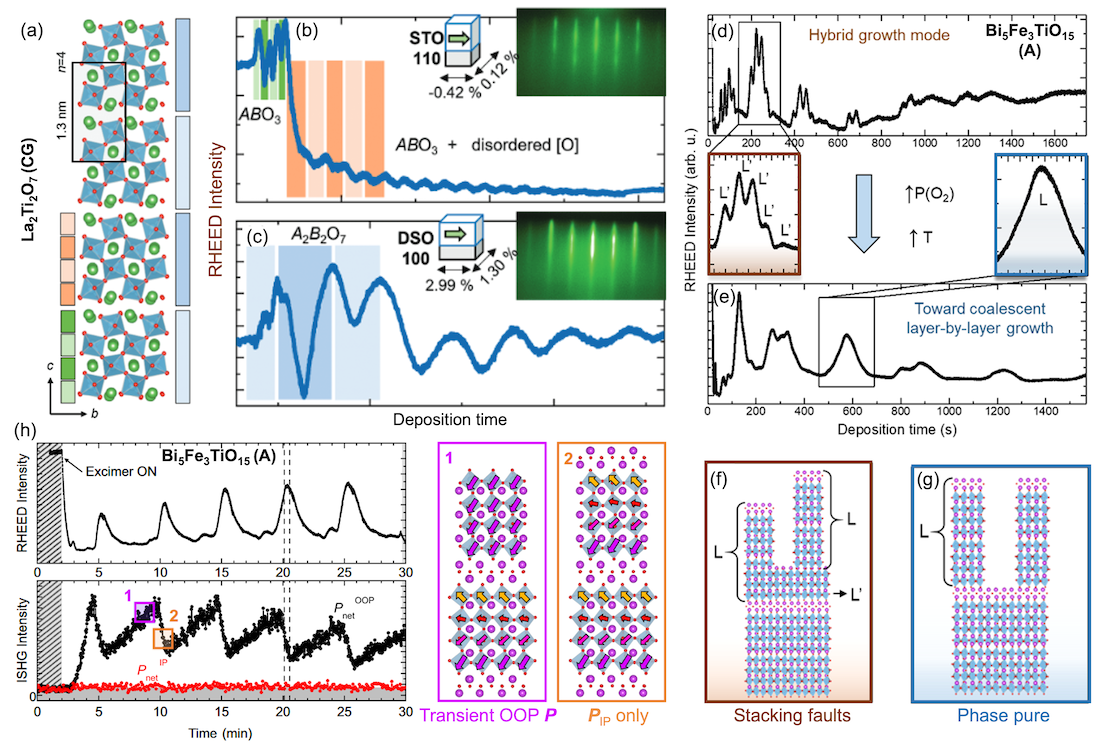}
  \end{adjustbox}
  \caption{\textbf{In-situ monitoring of structure and polarization evolution during the growth of layered ferroelectrics.} 
  (a–c) RHEED tracking of La\textsubscript{2}Ti\textsubscript{2}O\textsubscript{7} (Carpy--Galy) film (a) growth~\cite{gradauskaite2025}: (b) growth in perovskite-height layers on STO (110) leads to non-ferroelectric polymorphs, and (c) well-defined Carpy--Galy unit-cell growth on DSO (100) leads to the desired phase. Adapted with permission from~\citenum{gradauskaite2025}. Copyright 2025 Wiley‐VCH GmbH.  
  (d–g) RHEED monitoring of Bi\textsubscript{5}FeTi\textsubscript{3}O\textsubscript{15} (Aurivillius) growth~\cite{gradauskaite2021b}: (d) hybrid mode and (e) coalescent layer-by-layer mode. Schematics (f, g) illustrate that only the latter suppresses stacking faults and yields a phase-pure film. Adapted with permission from~\citenum{gradauskaite2021b}. Copyright 2021 American Chemical Society.  
  (h) Simultaneous monitoring of Bi\textsubscript{5}FeTi\textsubscript{3}O\textsubscript{15} growth with RHEED and ISHG signals~\cite{efe2025}. The oscillating ISHG signal reflects the transient emergence of out-of-plane polarization synchronized with structural layering (1,2). Adapted with permission from~\citenum{efe2025}. Copyright 2025 Springer Nature.}
  \label{fig:in_situ}
\end{figure}

\textbf{RHEED} is available for PLD, MBE, and sputtering, and it gives insights into how the crystallinity of the film develops during the growth and, in particular, on how the layering of perovskite planes and spacers is evolving.  Ex-situ X-ray reflectivity and atomic force microscopy (AFM) can be used in combination with RHEED monitoring to disentangle the structural origin of RHEED oscillations\cite{gradauskaite2020}. For instance, the growth of La$_2$Ti$_2$O$_7$ Carpy--Galy $n=4$ phase (Fig.\ \ref{fig:in_situ}a) on different substrates yields two distinct oscillation regimes\cite{gradauskaite2025}. If the RHEED intensity trace shows rapid oscillations (Fig.\ \ref{fig:in_situ}b), each corresponding to perovskite unit-cell thickness, the film is growing as kinetically favoured perovskite-type or other non-polar polymorphs. In order to achieve the CG phase in the as-grown films, one should aim to stabilize large-amplitude oscillations whose period matches the 1.3~nm height of an electroneutral $A_2B_2$O$_7$ block (Fig.\ \ref{fig:in_situ}c). Such a growth mode can be referred to as a ``coalescent'' layer-by-layer mode unique to layered oxides. Instead of adding individual perovskite planes sequentially, electroneutral (half-)unit-cell blocks nucleate separately and laterally merge into complete atomically flat layers. First demonstrated for the $n=4$ Aurivillius member using PLD\cite{gradauskaite2020}, the approach was later extended to $n=6$ and $n=8$ films\cite{gradauskaite2021b}. Signatures of such growth mode have also been reported in $n=5$ Aurivillius films grown by CVD\cite{keeney2023}.

Once the correct layered phase is stabilized, RHEED intensity oscillations can help to identify different $n$ intergrowths. For Bi$_5$FeTi$_3$O$_{15}$ $n=4$ films of the Aurivillius phase, ``hybrid'' growth modes can be observed, identified by the large unit-cell oscillations modulated by smaller ones (Fig.\ \ref{fig:in_situ}d). This signals that the perovskite planes are nucleating simultaneously with the large electroneutral layered unit cells. The result is an array of vertical stacking faults with higher $n$ (Fig.\ \ref{fig:in_situ}f). It was reported that raising the substrate temperature and oxygen pressure suppresses this mode (Fig.\ \ref{fig:in_situ}e) and leads back into the coalescent growth mode that helps to achieve fault-free films (Fig.\ \ref{fig:in_situ}g).

\textbf{\textit{In-situ} second-harmonic generation (ISHG)} offers a powerful, real-time optical tool to monitor symmetry breaking and polar order during film growth\cite{sarott2021a,deluca2017a}. A femtosecond laser is coupled into the growth chamber, and the second-harmonic signal emitted from the thin film is collected. In centrosymmetric crystals, SHG generation is forbidden symmetry. However, when inversion symmetry is broken by emergent polar order, SHG becomes allowed and a finite signal appears. By selecting specific input/output light polarization configurations, the SHG response can be made sensitive to either in-plane or out-of-plane (OOP) polarization components. The ISHG experiment tracking OOP polarization during the growth of the Aurivillius $n=4$ thin film\cite{efe2025} shows a sawtooth-like modulation closely related to structural RHEED oscillations (Fig.\ \ref{fig:in_situ}h). The SHG intensity gradually increases as each TiO\textsubscript{6}-based perovskite block is deposited and abruptly drops when the block is capped by the negatively charged Bi\textsubscript{2}O\textsubscript{2} spacer. This reveals that the transient OOP polarization arises from the electrostatic potential building up in the incompletely screened perovskite slab (Fig.\ \ref{fig:in_situ}h box 1). When the Bi\textsubscript{2}O\textsubscript{2} layer is deposited, it restores charge neutrality of the unit cell. As a result, the OOP component cancels out, and the unit cell reverts to the expected antipolar out-of-plane order, leaving only the in-plane polarization component (Fig.\ \ref{fig:in_situ}h box 2). This provides time-resolved evidence that Bi\textsubscript{2}O\textsubscript{2} layers govern the electrostatic landscape that stabilizes ferroelectricity in Aurivillius thin films.

\subsection{X-ray diffraction-based techniques}\label{sec:XRD}

XRD is typically the first method of choice for characterizing layered ferroelectric thin films, as it provides a full-volume probe of the crystalline structure. Unlike surface-sensitive techniques, XRD can detect parasitic phases or misoriented domains throughout the film thickness. Here we highlight some XRD-based approaches unique to layered ferroelectric thin films. 

\begin{figure}[htb!]
  \centering 
  \begin{adjustbox}{width=0.8\textwidth, center}
    \includegraphics[width=0.8\textwidth,clip, trim=5 5 3 5]{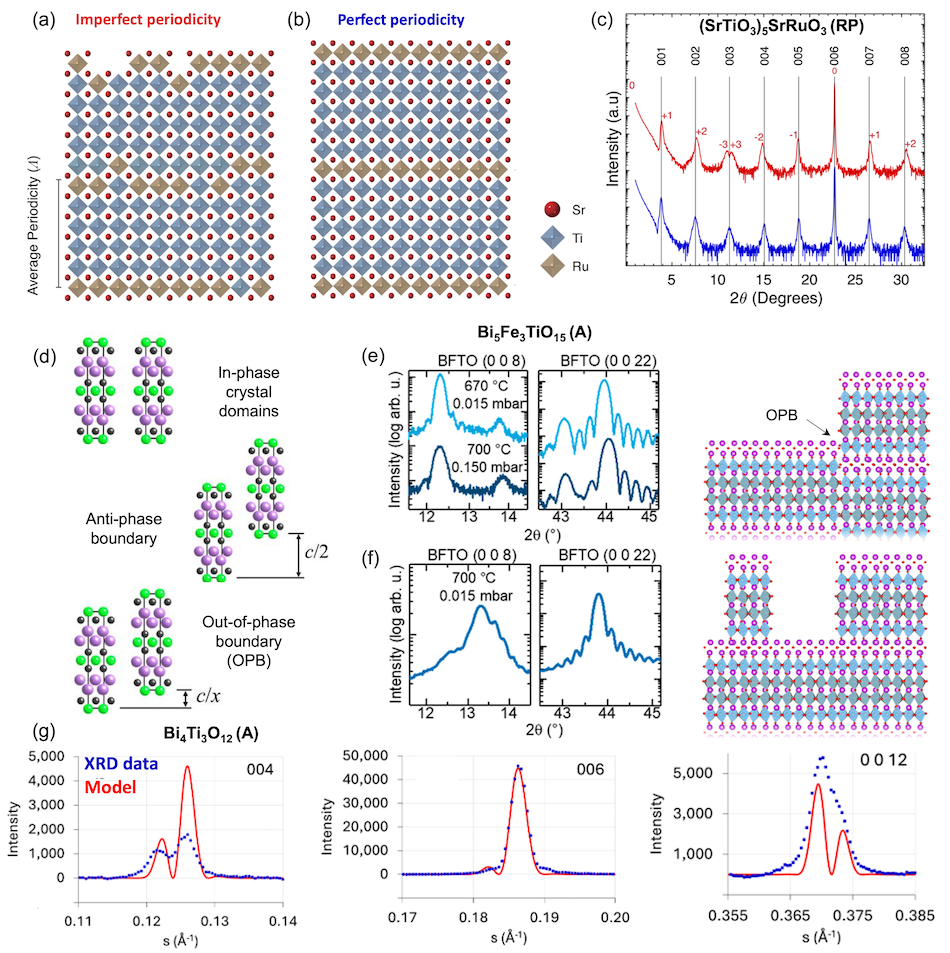}
  \end{adjustbox}
 \caption{\textbf{Diagnosing and modelling stacking faults in layered ferroelectric films by X-ray diffraction.} 
(a–c) Ruddlesden–Popper superlattices \((\mathrm{SrTiO}_{3})_{5}\mathrm{SrRuO}_{3}\) grown by MBE~\cite{barone2021c}: (a) ideal $n$=5 periodicity and (b) imperfect stacking with average periodicity of 0.9 unit-cell; (c) corresponding $\theta$–2$\theta$ scans, where deviations from the ideal sequence (red) broaden or even split the reflections (here in superlattice notation) compared with the phase-pure stack (blue). Reproduced with permission from~\citenum{barone2021c}. Copyright 2021 AIP Publishing.  
(d) Schematic illustrating in-phase unit cells, an anti-phase boundary where unit cells are shifted by exactly half a unit cell, and an out-of-phase boundary (OPB) where adjacent unit cells are misaligned by a fractional shift~\cite{zurbuchen2005}. Adapted with permission from~\citenum{zurbuchen2005}. Copyright 2005 Springer Nature.  
(e–f) OPB control in Bi\textsubscript{5}FeTi\textsubscript{3}O\textsubscript{15} (Aurivillius, \(n=4\)) films~\cite{gradauskaite2021a}: (e) splitting of the \((00\ell)\) peaks indicates OPB defects, (f) whereas improved growth conditions eliminate the splitting and yield defect-free films. Adapted with permission from~\citenum{gradauskaite2021a}. Copyright 2021 American Chemical Society.  
(g) Simulated XRD profiles from the OPB model (red) accurately reproduce the experimental Bi\textsubscript{4}Ti\textsubscript{3}O\textsubscript{12} thin-film data (blue) for the 0 0 4, 0 0 6 and 0 0 12 reflections~\cite{whatmore2025}. Reproduced with permission from~\citenum{whatmore2025}. Copyright 2025 International Union of Crystallography.}
  \label{fig:XRD}
\end{figure}

\textbf{Periodicity faults tracked by peak shape.}
Layered ferroelectric films often display imperfect periodicity (Fig.\ \ref{fig:XRD}a) despite sharing the same stoichiometry as perfectly periodic samples (Fig.\ \ref{fig:XRD}b), rendering many conventional characterization techniques insensitive to these structural variations. In $\theta$--2$\theta$ scans, periodicity faults broaden, skew, or split the XRD peaks (Fig.\ \ref{fig:XRD}c). Barone \textit{et al.}\cite{barone2021c} exploited this lineshape signature to iteratively tune flux ratios during MBE growth of Ruddlesden--Popper phases: successive runs sharpened the reflections and eventually enabled metastable $(\mathrm{SrTiO}_3)_{20}$-SrO $n=20$ stacks with perfect periodicity.

\textbf{Identification and modelling of out-of-phase boundaries.} When adjacent unit cells are perfectly aligned along the stacking axis, no defect is present. An anti-phase boundary forms when one unit cell is shifted vertically by exactly half a unit cell relative to its neighbor (Fig.~\ref{fig:XRD}d). If the structural shift differs from half a unit cell, the result is an out-of-phase boundary (OPB). OPBs are the most widespread defect in layered perovskite oxides and they most often nucleate at substrate steps, but can also arise from misfit dislocations or crystallographic shear when volatile species are lost during growth\cite{zurbuchen2007a}.

While observed in all layered ferroelectrics, OPBs are mostly investigated in Aurivillius films, where they give rise to XRD-peak splitting. For instance, in Bi$_5$FeTi$_3$O$_{15}$, the OPBs split specific $(00\ell)$ peaks, e.g.\ 0 0 8 and 0 0 22 \cite{gradauskaite2021b} (Fig.\ \ref{fig:XRD}e). Changing PLD growth conditions by raising temperature and oxygen partial pressure narrows or removes the peak split, indicating fewer OPBs (Fig.\ \ref{fig:XRD}f). Whatmore \textit{et al.}\cite{whatmore2025} introduced an analytical model that links physically accessible OPB parameters, such as the fractional–unit-cell shift across the OPB, the inclination angle at the film–substrate interface, and the average lateral spacing, to the resulting peak profile. With these inputs, the model accurately predicts which \((00\ell)\) reflections split and by how much, as confirmed by its agreement with the measured $\theta$–2$\theta$ pattern of a PLD-grown Bi\(_4\)Ti\(_3\)O\(_{12}\) film, see Figure \ref{fig:XRD}g. Another approach to quantitatively evaluate the nature and degree of stacking faults is to model diffuse X-ray scattering, as demonstrated for SrBi\textsubscript{2}Nb\textsubscript{2}O\textsubscript{9} thin films\cite{boulle2005a}. This allows for evaluation of the degree, scale, and evolution of planar defects in layered ferroelectric thin films.

\begin{figure}[b!]
  \centering 
  \begin{adjustbox}{width=0.57\textwidth, center}
    \includegraphics[width=0.57\textwidth,clip, trim=5 5 5 5]{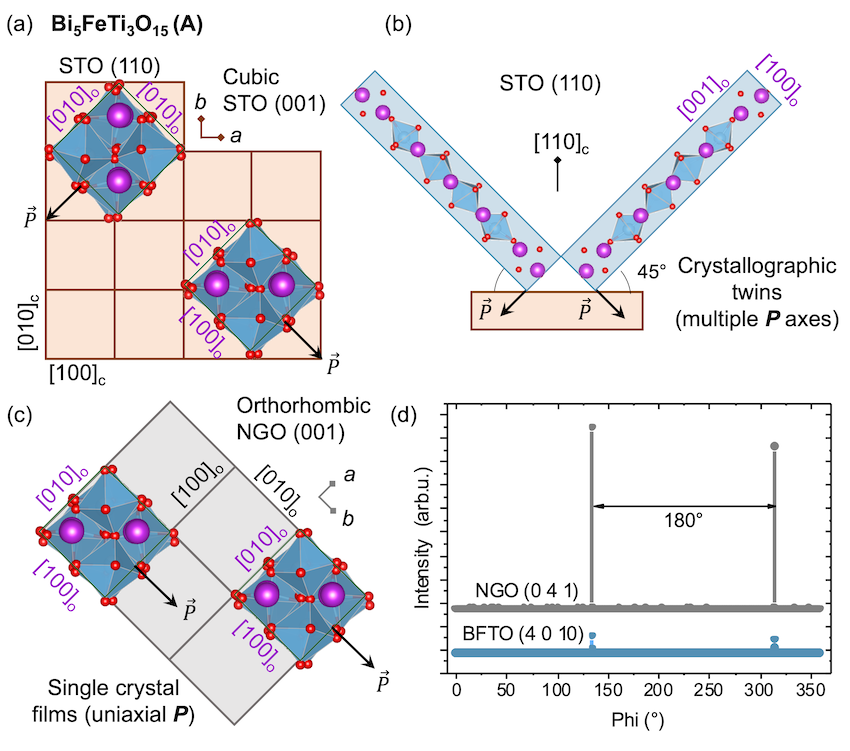}
  \end{adjustbox}
\caption{\textbf{Epitaxial matching of orthorhombic Aurivillius layered ferroelectrics.}
(a) In-plane view of an orthorhombic Bi\(_5\)FeTi\(_3\)O\(_{15}\) (BFTO) unit cell on a cubic SrTiO\(_3\)\,(001) substrate.  Lattice-vector mismatch enforces four in-plane polarization variants and generates 90° ferroelectric-polarization twins (can be regarded as ferroelastic domains\cite{wang2016c}).  
(b) When BFTO is grown on SrTiO\(_3\)\,(110): the polar axis tilts 45° from the surface normal, again producing twin domains and a finite out-of-plane polarization component.  
(c) On an orthorhombic, lattice-matched NdGaO\(_3\)\,(001) substrate, the BFTO \(a\)- and \(b\)-axes register one-to-one with the substrate, giving rise to a single-crystal film with no twinning and a uniaxial in-plane polarization.  
(d) X-ray diffraction $\phi$-scans confirm single-crystal films: the BFTO \((4\,0\,10)\) reflection (blue) coincides with the two-fold NdGaO\(_3\)\((0\,4\,1)\) peaks (grey), demonstrating complete epitaxial matching \cite{gradauskaite2020}. Adapted with permission from~\citenum{Gradauskaite2022thesis}.}
  \label{fig:epi_matching}
\end{figure}

Structural defects are often the link between the film microstructure and device performance in layered ferroelectrics. OPBs, although often labelled as “defects”, can play a beneficial role. In Aurivillius compounds, the celebrated fatigue-free switching\cite{dearaujo1995a} is attributed to the ability of the crystal to convert excess charged point defects into charge-neutral OPBs\cite{ding2001a,zurbuchen2007a}: they accommodate the compositional imbalance without degrading the ferroelectric response.  In Carpy–Galy phases, OPBs that appear during growth act as oxidation pathways that allow diffusion of extra oxygen, thereby stabilizing the layered structure starting from the perovskite phase\cite{seo2001c}.  Thus, rather than being purely detrimental, OPBs can serve as self-compensating or self-stabilizing features that aid the functional behavior of layered ferroelectric thin films.

\textbf{Crystallographic twinning from substrate mismatch.}
Unlike simple perovskite ferroelectrics such as PbTiO\(_3\) or BaTiO\(_3\), whose pseudocubic symmetry allows straightforward cube-on-cube epitaxy on common substrates like SrTiO\(_3\), epitaxy of layered ferroelectrics poses additional challenges. Their unit cells are typically orthorhombic or even monoclinic, with two different in-plane lattice parameters and a polar axis that lies within the plane of the film. As a result, direct growth on cubic substrates often leads to crystallographic twinning, where the film adopts multiple in-plane orientations to accommodate the symmetry mismatch. This not only affects film crystallinity but also suppresses net macroscopic polarization due to the presence of 90° ferroelastic domain variants.

Figure~\ref{fig:epi_matching} illustrates this issue using the Aurivillius Bi\(_5\)FeTi\(_3\)O\(_{15}\) (BFTO) films as an example. When grown on SrTiO\(_3\)\,(001) (Fig.\ \ref{fig:epi_matching}a), Aurivillius films form four rotational crystallographic and polarization variants due to in-plane symmetry mismatch\cite{wang2016c}. On SrTiO\(_3\)\ (110) (Fig.\ \ref{fig:epi_matching}b), crystallographic twins are also present with the polar axis tilting out of the plane, giving rise to an out-of-plane polarization component\cite{lettieri,ullah2019b}. However, when BFTO is grown on orthorhombic NdGaO\(_3\)\ (001), whose symmetry and lattice constants match those of BFTO\cite{gradauskaite2020,gradauskaite2021b,keeney2023} (Fig.\ \ref{fig:epi_matching}c), the film adopts a single-crystal structure with matched \(a\)- and \(b\)-axes and, therefore, a uniaxial in-plane polarization along a single substrate axis. This can be confirmed by X-ray $\phi$-scans (Fig.\ \ref{fig:epi_matching}c), where the BFTO \((4\,0\,10)\) peaks match the two-fold symmetry of the NdGaO\(_3\)\((0\,4\,1)\) reflection\cite{gradauskaite2020}. This example emphasizes that layered ferroelectrics require careful substrate selection to achieve both lattice and unit-cell symmetry matching.

\subsection{Scanning transmission electron microscopy} \label{sec:STEM}
STEM has become a key technique for imaging layered ferroelectric oxides at the atomic scale. It enables direct visualization of unit-cell layering, local polar displacements, mapping of ferroelectric domains, and the presence of structural defects such as OPBs or stacking faults. 

\begin{figure}[htb!]
  \centering 
  \begin{adjustbox}{width=0.85\textwidth, center}
    \includegraphics[width=0.85\textwidth,clip, trim=5 5 3 5]{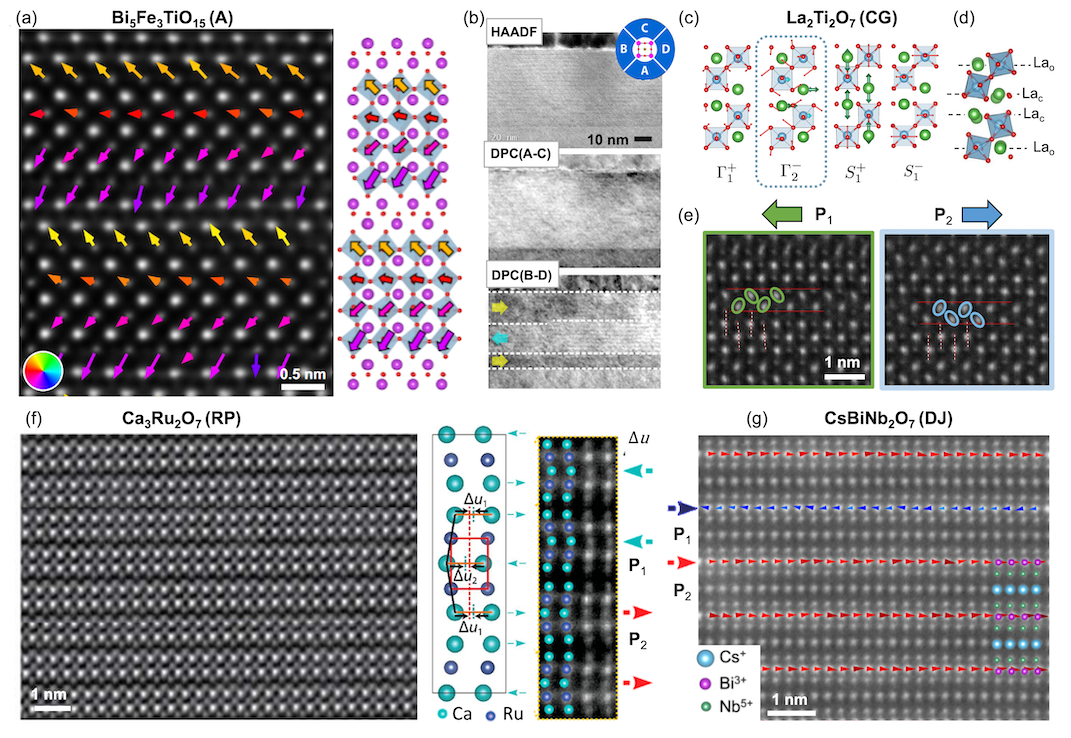}
  \end{adjustbox}
 \caption{\textbf{Atomic-scale mapping of polar displacements in layered ferroelectrics by STEM.}
(a,b) Bi$_5$FeTi$_3$O$_{15}$ (Aurivillius): (a) HAADF-STEM overlaid with dipole vectors reveals antipolar vertical shifts of TiO$_6$ octahedra toward neighbouring Bi$_2$O$_2$ layers with a net in-plane polarization\cite{efe2025}. Adapted with permission from~\citenum{efe2025}. Copyright 2025 Springer Nature. (b) Large-area HAADF and DPC images show out-of-plane (A–C) and in-plane (B–D) electric-field components. The latter gives access to large-scale imaging of buried in-plane domains\cite{campanini2019a}. Reproduced with permission from~\citenum{campanini2019a}. Copyright 2019 American Chemical Society.  
(c–e) La$_2$Ti$_2$O$_7$ (Carpy–Galy): symmetry-mode analysis (c) links distortion ($\Gamma_2^-$) to the ferroelectric $P2_1$ unit cell (d); HAADF-STEM (e) shows that lateral direction of La “teardrop’’ doublets (La\textsubscript{c} in (d)) are indicative of in-plane polarization direction\cite{gradauskaite2025}. Adapted with permission from~\citenum{gradauskaite2025}. Copyright 2025 Wiley-VCH GmbH.  
(f) Ca$_3$Ru$_2$O$_7$ (Ruddlesden–Popper): ADF-STEM image reveals bilayer structure with $A$-site lateral off-center displacements, with the direction alternating from layer to layer. The net $A$-site off-center displacement within each bilayer is $\Delta u = 2\Delta u_1 - \Delta u_2$ and is used to map in-plane polarization directions\cite{lei2018}. Adapted with permission from~\citenum{lei2018}. Copyright 2018 American Chemical Society.  
(g) CsBiNb$_2$O$_7$ (Dion–Jacobson): HAADF-STEM with Bi$^{3+}$ displacements relative to Cs$^{+}$ columns are used to map polarization, revealing unit-cell-wide 180$^{\circ}$ ferroelectric domains\cite{guo2021b}. Reproduced with permission from~\citenum{guo2021b}. Copyright 2021 American Physical Society.}
  \label{fig:stem}
\end{figure}

\textbf{Mapping of $B$-site off-centering in Aurivillius compounds.} In Aurivillius phases, the polar mode can be described by the simultaneous and antiparallel shifts of the Bi$_2$O$_2$ layers and the perovskite oxygen octahedra. While both the lateral displacement of Bi cations and the off-centering of $B$-site cations within oxygen octahedra can be imaged\cite{campanini2019a}, most studies focus on mapping the $B$-site cation displacements. This approach is sufficient to determine the polarization direction and follows established procedures from conventional perovskite ferroelectrics. As shown in Figure \ref{fig:stem}a, high-angle annular dark field (HAADF-) STEM imaging reveals antipolar stacking of $B$-site cation displacements along the out-of-plane direction and a net in-plane polarization within each perovskite block\cite{campanini2019a,gradauskaite2023,efe2025}. Additionally, differential phase contrast (DPC-) STEM has been proposed as an efficient technique for rapid, large-area mapping of buried in-plane domains in Aurivillius compounds\cite{campanini2019a} (Fig.~\ref{fig:stem}b). This technique relies on a four-quadrant segmented annular dark-field detector to resolve the projected electric field distribution.

\textbf{Octahedral rotation-driven polarization in Carpy–Galy ferroelectrics.}  
In Carpy–Galy compounds, ferroelectricity does not arise from simple $B$-site off-centering, but from a complex combination of distortions, predominantly oxygen octahedral rotations and lateral displacements of $A$-site cations. Symmetry-mode analysis for La$_2$Ti$_2$O$_7$, shown in Figure \ref{fig:stem}c, identifies distortion modes (the polar distortion is $\Gamma_2^-$) that give rise to the ferroelectric $P2_1$ cell (Fig.\ \ref{fig:stem}d). Within the ferroelectric unit cell, two types of $A$-site La cations can be distinguished: La ions closest to the oxygen spacers (La\textsubscript{c}) exhibit lateral, teardrop-like doublets, whereas the outer La ions (La\textsubscript{o}) remain nearly undistorted and appear spherical. These two distinct $A$-site cation environments create alternating zigzag-like features in HAADF-STEM images\cite{qiao2022b,gradauskaite2025}, with and without teardrop distortions, repeating every two atomic rows along the out-of-plane direction (Fig.~\ref{fig:stem}e). Notably, the direction of the La\textsubscript{c} doublet displacements correlates with the in-plane polarization direction, making them a reliable marker for mapping ferroelectric domains in Carpy–Galy films\cite{gradauskaite2025}.

\textbf{Mapping lateral $A$-site cation off-centering in Ruddlesden–Popper ferroelectrics.}  
The first atomically resolved STEM imaging of $A$-site cation displacements in a Ruddlesden–Popper ferroelectric was performed on Ca$_3$Ru$_2$O$_7$\cite{lei2018}. While technically a polar metal (Sec.\ \ref{sec:polar_metals}), this material exhibits polar displacements characteristic of hybrid improper ferroelectricity in Ruddlesden--Popper compounds. The annular dark field (ADF-) STEM image in Figure \ref{fig:stem}f reveals a bilayer structure with alternating lateral displacements of Ca across the film thickness. Lei \textit{et al.}\cite{lei2018} proposed an effective metric to quantify the net in-plane off-centering per bilayer, defined as $\Delta u = 2\Delta u_1 - \Delta u_2$, where $\Delta u_1$ and $\Delta u_2$ are the displacements in the two successive perovskite layers (Fig.\ \ref{fig:stem}f). This method enables direct mapping of in-plane polarization vectors. Later studies examined atomically-resolved microscopy of insulating hybrid improper ferroelectrics Li$_2$SrNb$_2$O$_7$\cite{uppuluri2019}, (Ca,Sr)$_3$Mn$_2$O$_7$\cite{miao2022}, and La$_2$SrSc$_2$O$_7$\cite{yi2024}.

\textbf{Unit-cell-wide domains in a Dion–Jacobson film.}  
In Dion–Jacobson films such as CsBiNb$_2$O$_7$, in-plane polarization can be mapped through lateral displacements of $A$-site cations with respect to the alkali spacers. In Figure \ref{fig:stem}g, Bi$^{3+}$ displacements relative to Cs$^{+}$ columns are used to reconstruct a polarization vector map, revealing 180$^\circ$ ferroelectric domains as narrow as one unit cell. These results, obtained on Dion--Jacobson films by Guo \textit{et al.}\cite{guo2021b}, demonstrate the weak interlayer coupling in these ferroelectric phases. Other than mapping ferroelectric domains, HAADF-STEM was also used to identify Ruddlesden--Popper intergrowths in the Dion--Jacobsons structure\cite{guo2021b}.

\subsection{Scanning probe microscopy} \label{sec:pfm}

Scanning probe microscopy provides a versatile, multimodal approach to characterize layered ferroelectric films. AFM can provide surface topography information and help identify phase intergrowths or signatures of layer-by-layer phase-pure growth \cite{gradauskaite2020,gradauskaite2021b,keeney2023}. However, piezoresponse force microscopy (PFM) provides the most important data by granting direct access to in-plane polarized ferroelectric domains. Finally, conductive AFM (c-AFM) can be used to image charged domain walls abundant in layered ferroelectrics\cite{oh2015c}.

\begin{figure}[htb!]
  \centering 
  \begin{adjustbox}{width=0.95\textwidth, center}
    \includegraphics[width=0.95\textwidth,clip, trim=3 3 3 3]{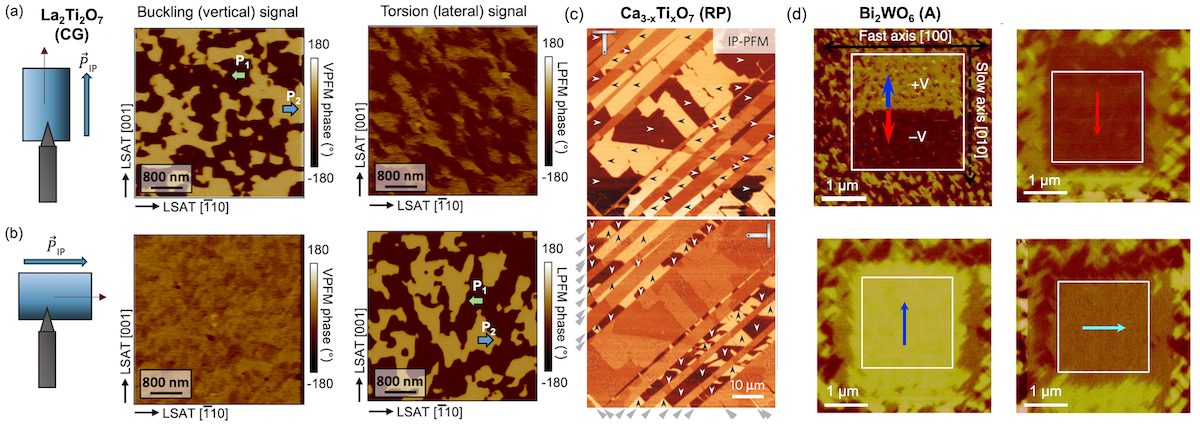}
  \end{adjustbox}
\caption{\textbf{Piezoresponse force microscopy (PFM) as a probe of in-plane polarization in layered ferroelectrics.}
(a,b) Vector PFM of a La$_2$Ti$_2$O$_7$ (Carpy--Galy) film\cite{gradauskaite2025}.  
With the cantilever parallel to the uniaxial polarization ($\mathbf P_1$) the contrast appears in the vertical (buckling) channel; a $90^\circ$ sample rotation transfers the same domain pattern to the lateral (torsion) channel and leaves the vertical response homogeneous, confirming purely in-plane polarization. Reproduced with permission from~\citenum{gradauskaite2025}. Copyright 2025 Wiley-VCH GmbH.  
(c) Lateral PFM of cleaved Ruddlesden–Popper Ca$_{2.46}$Sr$_{0.54}$Ti$_2$O$_7$ single crystals\cite{oh2015c}.  
The image contains two ferroelastic twin variants whose crystal axes, and hence in-plane polar axes, are rotated by 90$^\circ$ with respect to each other.  When the cantilever is aligned with the polar axis of one variant, that variant shows strong PFM contrast while the other appears nearly featureless.  Rotating the sample by 90$^\circ$ swaps the situation, so the previously unresolved crystallographic twin now reveals its $180^\circ$ ferroelectric domains (white and black arrows) and the contrast within the first variant vanishes. Reproduced with permission from~\citenum{oh2015c}. Copyright 2015 Springer Nature.  
(d) Trailing-field-based switching of a Bi$_2$WO$_6$ thin film\cite{wang2016c}. By scanning with $\pm10$–15\,V applied to the AFM tip, an in-plane trailing electric field is generated by the AFM probe. This field reorients the ferroelectric polarization along the slow-scanning direction, enabling the deterministic writing of $90^\circ$-rotated in-plane domains. Adapted with permission from~\citenum{wang2016c}. Copyright 2016 Springer Nature.}
  \label{fig:pfm}
\end{figure}

\textbf{Vector PFM of uniaxial in-plane polarization.}  
In single-crystal layered ferroelectric films, the polar axis lies entirely in-plane and is uniaxial\cite{gradauskaite2020,gradauskaite2025}.  As shown in Fig.~\ref{fig:pfm}(a,b), aligning the AFM cantilever parallel to this axis produces strong contrast in the vertical (buckling) channel, while a 90° sample rotation shifts the same domain pattern into the lateral (torsion) channel.  These complementary measurements upon sample rotation confirm pure in-plane polarization along a single axis without an out-of-plane component\cite{gradauskaite2025} and therefore single-crystallinity of the layered ferroelectric film.  

\textbf{Resolving ferroelastic twin variants.}  
Layered perovskites on mismatched substrates (or single crystals) form ferroelastic twins with the polar axis rotated by 90°.  In cleaved Ruddlesden--Popper Ca\(_{2.46}\)Sr\(_{0.54}\)Ti\(_2\)O\(_7\) crystals, orienting the cantilever perpendicular to one variant’s polar axis yields strong PFM contrast in that twin while the orthogonal variant shows no contrast (Fig.\ \ref{fig:pfm}c). A 90° sample rotation then reveals the previously hidden $180^\circ$ in-plane domains in the second twin, enabling complete mapping of polarization in both crystallographic variants\cite{oh2015c}.  

\textbf{Trailing-field switching of in-plane domains.}  
Even without a bottom electrode, deterministic domain writing is possible via the trailing electric field of an AFM tip\cite{gradauskaite2020, Strkalj2021}. As illustrated in Figure \ref{fig:pfm}d, scanning with a DC bias of \(\pm 10\)–15 V generates an in-plane field beneath the tip that aligns the ferroelectric polarization along the slow-scanning direction.  This method allows on-demand writing of $90^\circ$-rotated in-plane ferroelastic domains in Aurivillius films such as Bi\(_2\)WO\(_6\)\cite{wang2016c}.  

\textbf{c-AFM for mapping conduction at charged domain walls.} 
In oxygen-deficient Ruddlesden--Popper Ca\(_{2.44}\)Sr\(_{0.56}\)Ti\(_2\)O\(_{7-\delta}\) crystals, c-AFM was employed for mapping charged domain walls in conjunction with lateral PFM imaging\cite{oh2015c}. The polarity-dependent conductivity was reported: positively charged head-to-head walls accumulate \(n\)-type carriers and conduct well, whereas negatively charged tail-to-tail walls repel carriers and remain insulating\cite{oh2015c}.  Charged domain walls, rare in perovskite ferroelectrics due to their high electrostatic cost, emerge naturally in layered ferroelectrics and can even be engineered through structural defects in thin films (Sec.\ \ref{sec:OPB_domains}).

\subsection{Other complementary techniques}  
Several other methods help complete the characterization of layered ferroelectrics. Ferroelectric switching measurements, performed with interdigitated top electrodes to switch the in-plane polarization\cite{reader2020}, are essential for assessing it quantitatively and testing its endurance\cite{asaki2020b,gradauskaite2020,gradauskaite2025}. To isolate the intrinsic ferroelectric response from leakage contributions, the Positive-Up Negative-Down (PUND) method is frequently employed. Optical second-harmonic generation (SHG) enables domain and domain-wall imaging\cite{weber2022}; additionally, temperature sweeps of the integrated SHG signal yield phase transition temperatures\cite{zemp2024}.  
Neutron diffraction techniques applied to single crystals of layered ferroelectrics are indispensable for resolving subtle distortions such as oxygen octahedral rotations\cite{hervoches2002a,snedden2003d, snedden2003} or incommensurate structural modulations\cite{howiesonIncommensurateCommensurateTransition2020} that are often invisible in conventional XRD.  
Finally, dielectric spectroscopy\cite{chen2015, asaki2020b} offers means of probing $T_{\mathrm{C}}$ and possible relaxor behavior. 

\section{Uncovering novel functionalities in epitaxial layered ferroelectrics}

Layered ferroelectrics were first explored mainly as polycrystalline ceramics or bulk single crystals. Only in the past decade the focus has shifted toward thin films (often epitaxial), where advances in growth and characterization now allow us to probe their behavior at the unit-cell thickness scale. High-quality epitaxial films with well-defined polarization orientations have revealed not only the retention of bulk-like functionalities but also unexpected new phenomena that do not occur in conventional perovskites. In the following, we first discuss how epitaxy enables thin films to retain and in some cases even enhance the functional properties known for bulk layered ferroelectrics (Sec.\ \ref{sec:enhanced}). We then highlight the absence of a critical thickness for ferroelectricity in in-plane-polarized films (Sec.\ \ref{sec:critical_thickness}), and describe how the periodic electrostatics of their unit cells in combination with structural defects can stabilize unconventional polar textures such as charged domain walls and vortices (Sec.\ \ref{sec:OPB_domains}). Finally, we outline how the structural compatibility of layered ferroelectrics with perovskites enables new types of hybrid heterostructures and composites (Sec.\ \ref{sec:perovskites}).

\subsection{Retaining and enhancing bulk properties in layered thin films} \label{sec:enhanced}

Many of the functional properties originally demonstrated in single crystals or polycrystalline ceramics of layered ferroelectrics have successfully translated to the thin-film geometry, retaining their stability down to just a few unit cells. For instance, the Aurivillius family was historically used for fatigue-resistant ferroelectric capacitors in the 1990s\cite{dearaujo1995a}. The exceptional endurance against ferroelectric fatigue has now been confirmed in ultra-thin films of just 2.5 unit cells, where the ferroelectric response persists for over 10$^{10}$ cycles with a significant in-plane polarization of 16.5 \textmu C/cm$^2 $\cite{gradauskaite2020}.

In certain cases, the thin-film form not only preserves but significantly enhances the functional properties previously reported in bulk single crystals. For instance, epitaxial stabilization of Carpy–Galy La$_2$Ti$_2$O$_7$ on SrTiO$_3$ (110) substrates enabled recording a polarization value of 18.3 \textmu C/cm$^2$ \cite{gradauskaite2025}, which is nearly four times higher than previously reported in bulk crystals\cite{nanamatsuNewFerroelectricLa2Ti2o71974}. Simultaneously, the coercive field was substantially reduced, facilitating easier polarization switching. These improvements are attributed to epitaxial control over film orientation, which minimizes crystallographic intergrowths and enforces uniform unit-cell stacking. Such findings suggest that other overlooked or underestimated compounds of layered ferroelectrics could similarly reveal enhanced functionality when synthesized as high-quality epitaxial thin films. 

Modern thin films of layered ferroelectrics are also considered for new application pathways. For example, the same fatigue-resistant response that originally made Aurivillius compounds attractive for ferroelectric memory can be re-engineered and optimized for energy storage performance\cite{pan2020b}. Here, the materials are leveraged not for their remanent polarization but for their large and recoverable dielectric response.

\subsection{Absence of critical thickness for ferroelectricity}\label{sec:critical_thickness}

\begin{figure}[htb!]
  \centering 
  \begin{adjustbox}{width=0.87\textwidth, center}
    \includegraphics[width=0.87\textwidth,clip, trim=5 1 3 5]{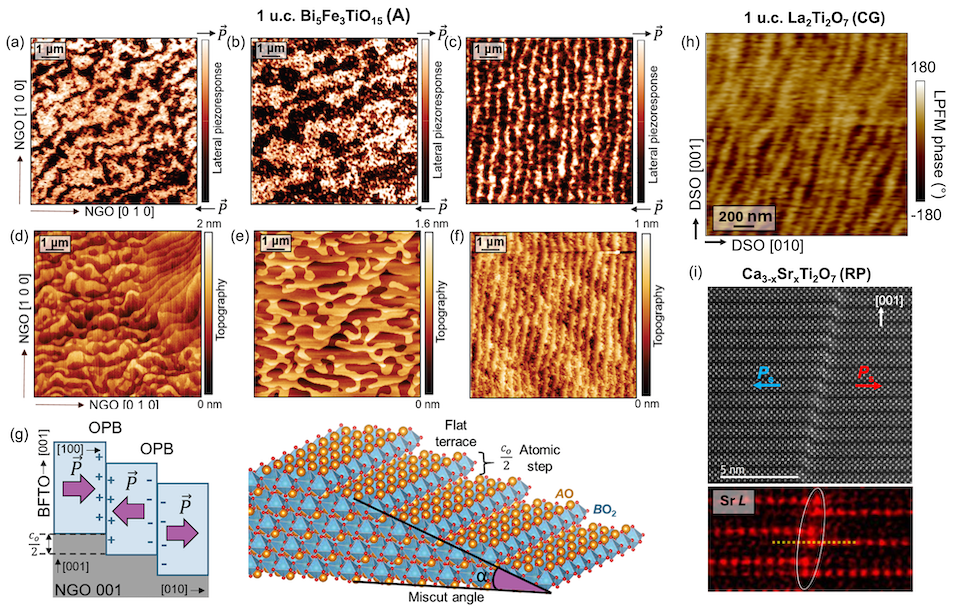}
  \end{adjustbox}
\caption{\textbf{Out-of-phase boundaries (OPBs) as nucleation centres for charged ferroelectric domain walls in layered films.}
(a) Irregular striped domain network in a 1\,u.c.\ Bi$_5$FeTi$_3$O$_{15}$ (BFTO) film corresponding to (d) the topography of a small-miscut (0.02$^\circ$) untreated substrate.
(b,e) A wider-domain pattern in 1\,u.c.\ BFTO layer is achieved after substrate annealing that produces wider terraces.  
(c,f) Highly ordered stripe domains in 1\,u.c.\ BFTO can be produced by a larger miscut (0.05$^\circ$) substrate with narrowly spaced straight steps.  
(g) Schematics illustrating how every single substrate step in NdGaO$_3$ (001) generates a steric OPB in a layered BFTO
film\cite{gradauskaite2022}; It acts as a nucleation site of charged domain walls due to the altered electrostatic and bonding environment and susceptibility to nonstoichiometric defects. Adapted with permission from~\citenum{gradauskaite2022}. Copyright 2022 American Chemical Society.  
(h) In‐plane PFM image of a 1 u.c.-thick La$_2$Ti$_2$O$_7$ film\cite{gradauskaite2025} showing uniform stripe domains, corroborating OPB-driven domain-wall formation. Reproduced with permission from~\citenum{gradauskaite2025}. Copyright 2025 Wiley‐VCH GmbH.  
(i) HAADF‐STEM and energy‐dispersive X-ray spectroscopy of a charged tail-to-tail domain wall coinciding with an OPB in Ca$_{3-x}$Sr$_x$Ti$_2$O$_7$\cite{nakajima2021a} crystals.  Sr segregation at the OPB suggests the mechanism of charge accumulation at such boundaries that can stabilize charged domain walls. Adapted with permission from~\citenum{nakajima2021a}. Copyright 2021 Springer Nature.}
  \label{fig:OPB_domains}
\end{figure}

In‐plane-polarized ferroelectric films should, in principle, exhibit no critical thickness.  Because the polarization lies parallel to the film surface, no bound charge accumulates at the film surfaces. Therefore, the depolarizing fields that destabilize out‐of‐plane ferroelectrics below four unit cells\cite{Junquera2003,deluca2017a} are absent.  Only very recently, however, thin film growth with unit-cell-thickness precision enabled by RHEED (Sec.~\ref{sec:in_situ_growth}) made it possible to test this experimentally.
Clear stripe-domain contrast in lateral PFM was observed starting from just 0.5 unit cell in Aurivillius films\cite{gradauskaite2020,gradauskaite2021b,gradauskaite2022} (Fig.\ \ref{fig:OPB_domains}a–c) and in 1 unit-cell-thick Carpy–Galy films\cite{gradauskaite2025} (Fig.\ \ref{fig:OPB_domains}h), both grown by PLD. Similar results were obtained for Aurivillius films synthesized by CVD, both in near-unit-cell-thick films\cite{keeney2020d} and in thicker films that were progressively thinned to a few nanometers using an AFM tip to locally remove material\cite{keeney2023}. The absence of a critical thickness was further confirmed in exfoliated sub-unit-cell Aurivillius flakes\cite{keeney2020c}. Dion–Jacobson ferroelectric nanosheets prepared by mechanical exfoliation also exhibit a thickness-independent piezoresponse\cite{shimada2025}, reinforcing the generality of this behavior across layered ferroelectrics.

The convergence of evidence from Aurivillius, Carpy–Galy, and Dion–Jacobson systems, prepared by PLD, CVD, and mechanical exfoliation of single crystals, confirms that the absence of a depolarizing field in in-plane-polarized layered ferroelectrics is sufficient to eliminate the classical critical thickness. This positions layered ferroelectrics as attractive candidates for ultrathin, energy-efficient memories, sensors, and actuators.

\subsection{Non-trivial polar textures: charged domain walls and vortices} \label{sec:OPB_domains}.

Layered ferroelectrics can stabilize polarization configurations that are typically energetically unfavorable in conventional systems. These non-trivial polar textures, such as charged domain walls and vortices, arise due to structural and electrostatic discontinuities introduced by OPBs. As a result, configurations that would be suppressed in standard ferroelectrics can emerge and persist in layered systems. We discuss some examples below.

\textbf{Charged domain walls nucleated at out-of-phase boundaries.} In addition to demonstrating the absence of critical thickness for ferroelectricity in layered in-plane ferroelectrics, PFM performed on unit-cell-thick Aurivillius films revealed unexpected stripe-like ferroelectric domain patterns\cite{gradauskaite2020} (Fig.\ \ref{fig:OPB_domains}a-c). Notably, the domain walls run perpendicular to the uniaxial in-plane polarization of the film, forming a periodic array of alternating charged head-to-head and tail-to-tail domain walls (Fig.\ \ref{fig:OPB_domains}g). Charged domain walls are promising for domain-wall nanoelectronics, as they exhibit conduction properties distinct from the surrounding bulk\cite{Catalan2012,Meier2012,Nataf2020}. However, they are typically energetically unfavorable in proper ferroelectrics, implying an unconventional formation mechanism in layered ferroelectrics.

By comparing domain patterns imaged by PFM (Fig.\ \ref{fig:OPB_domains}a–c) with the corresponding substrate topography (Fig.\ \ref{fig:OPB_domains}d–f), a one-to-one correlation was established between substrate steps and domain walls\cite{gradauskaite2022}. Density-functional theory and phase-field simulations attribute this to the formation of OPBs\cite{gradauskaite2022}: when the layered unit cell crosses a step terrace of the substrate, it shifts vertically by a fraction of its height along the out-of-plane direction (Fig.\ \ref{fig:OPB_domains}g). This structural discontinuity alters the local electrostatic boundary conditions and is prone to non-stoichiometries\cite{zurbuchen2007a,gradauskaite2022}, which induce and stabilize charged domain walls. Similar OPB-pinned stripes were later observed in other compositions of Aurivillius films\cite{keeney2020d} as well as in 1-u.c.\ La\(_2\)Ti\(_2\)O\(_7\) (Carpy–Galy) films\cite{gradauskaite2025} (Fig.~\ref{fig:OPB_domains}h), again mirroring the underlying substrate terrace morphology. OPBs that host charged domain walls were also observed in Ca\(_{3-x}\)Sr\(_x\)Ti\(_2\)O\(_7\) crystal of Ruddlesden–Popper phase, in which  HAADF-STEM revealed a charged tail-to-tail domain wall coincident with an OPB (Fig.~\ref{fig:OPB_domains}i)\cite{nakajima2021a}. Energy-dispersive X-ray spectroscopy showed local Sr enrichment at the OPB, suggesting the accumulation of ionic charge that compensates the bound negative polarization charge of the wall.  

The recurring motif across three different families of layered ferroelectrics suggests that such OPB-induced charged domain walls are universal for in-plane-polarized layered ferroelectrics. This could be explored for precisely engineered locations of charged domain walls and their density by tailoring substrates on which the films grow.

\begin{figure}[htb!]
  \centering 
  \begin{adjustbox}{width=0.7\textwidth, center}
    \includegraphics[width=0.7\textwidth,clip, trim=5 1 3 5]{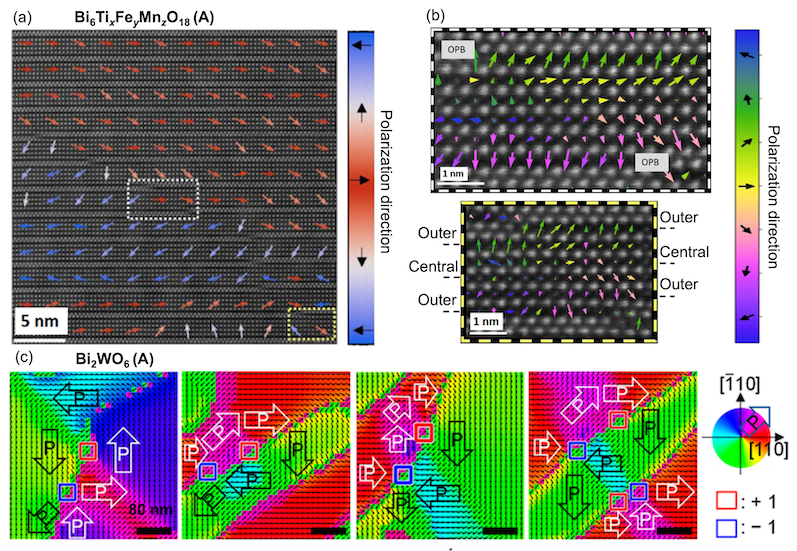}
  \end{adjustbox}
\caption{\textbf{Polar‐vortex textures stabilized in layered ferroelectric thin films.}
  (a) Polarization map of a Bi$_{6}$Ti$_{x}$Fe$_y$Mn$_z$O$_{18}$ Aurivillius film\cite{moore2022b}: two vortices (dashed boxes) emerge where tail-to-tail domain walls intersect OPBs. Adapted with permission from~\citenum{moore2022b}. Copyright 2022 American Chemical Society.  
  (b) Zoomed-in images of the white and yellow boxes in (a).  The OPB breaks the lattice periodicity, generates a strain/electrostatic discontinuity, producing a polarization vortex.  
  (c) Angle-resolved PFM analysis of an epitaxial Bi$_2$WO$_6$ film\cite{kwon2023}.  
  Six lateral-PFM phase maps, acquired at different cantilever orientations, are combined to reconstruct local polarization vectors, revealing flux-closure vortex structures (red and blue boxes denote winding numbers $+1$ and $-1$). The vortices are stabilized by a “hidden” $\langle100\rangle$-polarized phase that bridges the four dominant $\langle110\rangle$ variants. Reproduced with permission from~\citenum{kwon2023}. Copyright 2023 American Chemical Society.}
  \label{fig:vortex}
\end{figure}

\textbf{Polar vortices in films of layered ferroelectrics.} Moore \textit{et al.}\cite{moore2022b} investigated multiferroic\cite{keeney2013a, faraz2017a,keeney2017a} Aurivillius Bi$_{6}$Ti$_{x}$Fe$_{y}$Mn$_{z}$O$_{18}$ $n=5$ films and discovered nanoscale flux-closure vortices. At OPBs, the film is subject to both a strain gradient and a discontinuity in electrostatic boundary conditions. When a charged domain wall coincides with an OPB—as revealed by atomically resolved STEM polarization vector mapping (Fig.\ \ref{fig:vortex}a), polarization can be forced into a continuous rotation, producing a vortex (Fig.~\ref{fig:vortex}b). Such a vortex should be charged in the same way as tail-to-tail domain walls are, but is expected to have the charge more localized at the vortex core, further increasing local conductivity. For the vortex to form, the lateral distance between successive OPBs must be similar to the 180\textdegree{} charged domain-wall width, which is around 5--8 perovskite cells\cite{moore2022b}. Because OPB density can be tuned via substrate miscut\cite{gradauskaite2022} or growth stoichiometry\cite{zurbuchen2007a}, the vortices in Aurivillius films can, in principle, also be engineered.

Vortex-like polar textures were also uncovered in epitaxial Aurivillius Bi$_2$WO$_6$ films. Kwon \textit{et al.}\,\cite{kwon2023} applied a high-resolution angle-resolved lateral PFM  protocol, acquiring six lateral PFM maps at different cantilever azimuths and reconstructing the full local piezoresponse vector map.  This analytical approach uncovered a “hidden” polar phase that links the four conventional polarization variants that appear in flux-closure vortex domain patterns.  The bridge phase enables smooth $90^{\circ}$ rotations and organizes polarization variants into flux-closure vortices with a winding number $+1$ or $-1$ (Fig.~\ref{fig:vortex}c).

\subsection{Epitaxial integration of layered ferroelectrics with perovskites}\label{sec:perovskites}

The structural similarity between layered ferroelectrics and conventional perovskites enables their epitaxial integration into heterostructures. Such compatibility opens avenues for engineering artificial materials that combine in-plane-polarized layered ferroelectrics with conventional perovskites that can be metallic, magnetic, or out-of-plane-polarized. In this way, one can create novel interfacial phenomena and access functional properties beyond those of the individual phases.

\begin{figure}[htb!]
  \centering 
  \begin{adjustbox}{width=0.95\textwidth, center}
    \includegraphics[width=0.95\textwidth,clip, trim=5 5 5 5]{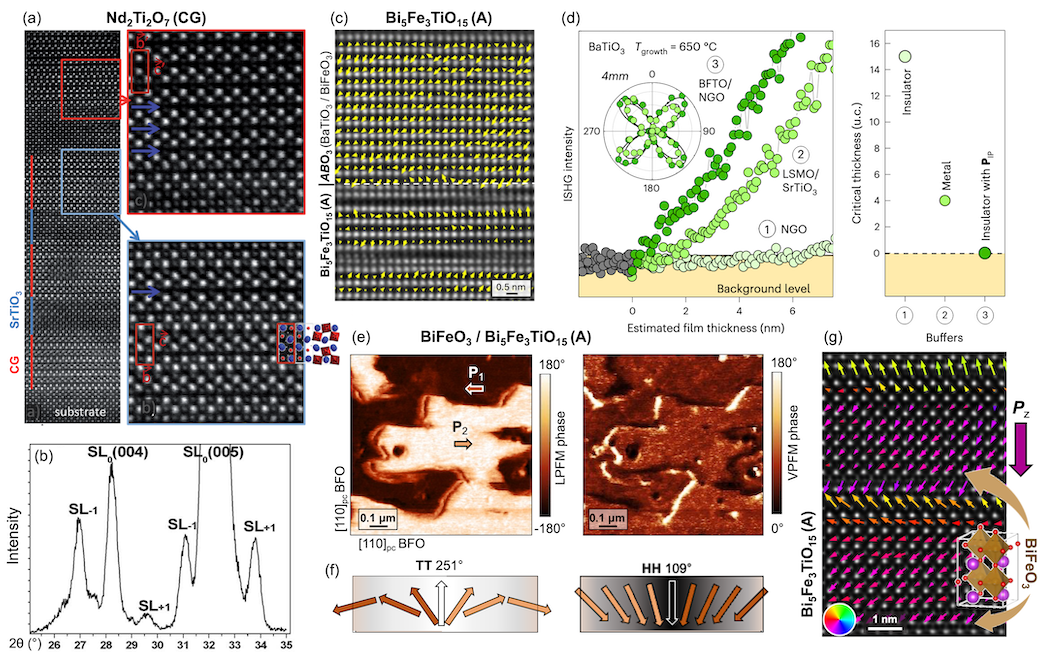}
  \end{adjustbox}
  \caption{\textbf{Epitaxial integration of layered ferroelectrics with perovskites.}
  (a,b) Coherent stacking of Nd\(_2\)Ti\(_2\)O\(_7\) (Carpy–Galy) and SrTiO\(_3\) perovskite blocks in a \(\bigl[\)Nd\(_2\)Ti\(_2\)O\(_7\)\(_4\)/SrTiO\(_3\)\(_4\)]\(_{10}\) superlattice\cite{carlier2018a}. (a) The STEM image shows alternating Carpy--Galy (red markers) and STO perovskite (blue markers) layers.  Zoom-in into the blue-framed region reveals a single kinetically-favored metastable $\gamma$ polymorph\cite{havelia2009c,bayart2013a} stacking fault (arrow), which repeats periodically at higher order (red frame). (b) Superlattice satellite peaks were observed for the two brightest Carpy--Galy XRD reflections. Reprinted with permission from~\citenum{carlier2018a}. Copyright 2018 Royal Society of Chemistry.  
  (c-f) Use of in-plane-polarized Aurivillius Bi\(_5\)FeTi\(_3\)O\(_{15}\) (BFTO) as a buffer for perovskite BaTiO\(_3\) and BiFeO\(_3\)\cite{gradauskaite2023}.  
  (c) HAADF–STEM and $B$-site displacement maps at the BiFeO\(_3\)/BFTO interface demonstrate unit-cell-sharp coherence and immediate onset of ferroelectric polarization with no critical thickness.  
  (d) ISHG intensity during BaTiO\(_3\) growth on (1) bare substrate, (2) metallic electrode, and (3) in-plane-polarized BFTO buffer, showing that only the BFTO buffer eliminates the usual critical thickness for ferroelectricity.  (e) Lateral and vertical PFM maps of BiFeO\(_3\) films grown on BFTO reveal two in-plane polarization variants and a uniform downward out-of-plane polarization, with some discontinuities appearing at the domain-wall contours. These are consistent with homochiral ferroelectric domain walls with BiFeO\(_3\)\cite{gradauskaite2023} with associated polarization profiles schematically shown in (f). Adapted with permission from~\citenum{gradauskaite2023}. Copyright 2023 Springer Nature.  
  (g) HAADF–STEM overlaid with polarization vectors shows a hybrid order in BiFeO\(_3\)/BFTO composite film, whereby out-of-plane dipoles no longer cancel out in the out-of-plane direction between the fluorite spacers. This yields a ferrielectric-like state with a net out-of-plane polarization in addition to the in-plane polarization\cite{efe2025}. Adapted with permission from~\citenum{efe2025}. Copyright 2025 Springer Nature.}
  \label{fig:perovskite}
\end{figure}

\textbf{Carpy–Galy and perovskite superlattices.}  
A first demonstration of coherent epitaxy between layered and simple perovskite oxides was achieved in Nd\(_2\)Ti\(_2\)O\(_7\)/SrTiO\(_3\) superlattices grown by PLD\cite{carlier2018a}. Here, the Carpy–Galy $n=4$ ferroelectric is epitaxially combined with the cubic SrTiO\textsubscript{3} perovskite layer\cite{carlier2018a}. HAADF-STEM reveals a coherently strained superlattice (Fig.\ \ref{fig:perovskite}a), that preserves a Carpy--Galy layering (blue box shows a periodic pattern of $A$-cation zigzags, refer to Section \ref{sec:STEM}). A kinetically-favored metastable $\gamma$ polymorph\cite{havelia2009c,bayart2013a} appears as a stacking fault in the layers (marked by a blue arrow) and becomes more frequent with increasing thickness, see the red inset. Superlattice satellites in XRD (Fig.\ \ref{fig:perovskite}b) confirm a well-defined repeat distance and a high quality of such hybrid superlattices between Carpy--Galy and perovskite phases.  Local piezoresponse spectroscopy showed that the Nd\(_2\)Ti\(_2\)O\(_7\) ferroelectric response is attenuated in the superlattice, while current–voltage mapping revealed conductive channels attributed to oxygen-vacancy tunneling across interfaces\cite{carlier2018a}. 

\textbf{In-plane-polarized Aurivillius buffer for out-of-plane-polarized perovskite ferroelectrics.}  Out-of-plane ferroelectric thin films typically exhibit a critical thickness of at least 4 unit cells, below which ferroelectricity is suppressed by depolarizing fields\cite{Junquera2003,deluca2017a}. By integrating an in-plane-polarized Aurivillius buffer, which remains ferroelectric down to sub-unit-cell thickness\cite{keeney2020d,gradauskaite2020} (Sec.\ \ref{sec:critical_thickness}), it becomes possible to enforce polarization continuity across the interface and stabilize out-of-plane polarization from the very first unit cell in BaTiO\(_3\) and BiFeO\textsubscript{3} (BFO) films\cite{gradauskaite2023}. Such heterostructures show an atomically sharp interface between the 1-u.c.-thick Aurivillius Bi\(_5\)FeTi\(_3\)O\(_{15}\) (BFTO) buffer layer and the perovskite ferroelectric film, see Figure \ref{fig:perovskite}c. ISHG measurements reveal the out-of-plane polarization onset from the first unit cell when these ferroelectrics are grown on the layered ferroelectric buffer, which is a significant improvement when compared to the critical thickness of 4 and 15 unit cells that is measured for the films grown on a metallic buffer and directly on the substrate, respectively. 
 
When multiferroic BFO is deposited onto a BFTO buffer layer, the in-plane-polarized buffer of the Aurivillius phase not only removes the critical thickness for ferroelectricity, but also changes its allowed ferroelectric domain and domain-wall variants\cite{gradauskaite2023,gradauskaite2025c}. The BFO film grown on BFTO exhibits two in-plane polarization variants and a uniform downward out-of-plane polarization (Fig.\ \ref{fig:perovskite}e). At the domain walls, however, vertical polarization exhibits local discontinuities, implying a continuous rotation of the polarization vector. Specifically, tail-to-tail (TT) walls rotate upward, forming 251$^\circ$ domain walls, while head-to-head (HH) walls rotate downward across 109$^\circ$, leading to a defined polarization rotation sense at all walls (Fig.\ \ref{fig:perovskite}f). This homochiral domain-wall character is consistent with a ferroelectric analogue to the Dzyaloshinskii–Moriya interaction (DMI)\cite{Zhao2021} and is induced by the Aurivillius buffer layer\cite{gradauskaite2023,gradauskaite2025c}.

\textbf{Composite Aurivillius–perovskite films.}  
Going beyond epitaxial heterostructures between perovskites and layered ferroelectrics, it was demonstrated that it is even possible to embed perovskite layers directly into the Aurivillius crystal framework using ISHG and RHEED monitoring during the deposition\cite{efe2025}. In this way, one can create artificial composite systems with additional perovskite layers of choice sandwiched between the Bi-rich fluorite-like spacers. Figure \ref{fig:perovskite}(g) shows a composite film in which BFO unit cells are included within the Aurivillius  BFTO $n=4$ matrix\cite{efe2025}.  Atomically sharp Bi\(_2\)O\(_2\) layers are preserved, while allowing Ti and Fe to intermix across perovskite blocks. Polar vector mapping reveals a ferrielectric-like state along the out-of-plane direction: $B$-site dipoles now form a net out-of-plane polarization component. Standard Aurivillius compounds with $n$=even exhibit no net out-of-plane polarization, see Figure \ref{fig:stem}a. However, by inserting additional BFO layers into the framework, the electrostatic balance between the spacer layers is disrupted, and an out-of-plane component of polarization, coexisting with the native in-plane polarization of the layered compound, is stabilized. The inclusion of BFO also controls the magnetic functionality of the composite, as its incorporation induces antiferromagnetic order\cite{efe2025}. All this demonstrates that composite films retain the functionality of their Aurivillius and perovskite constituents, offering an exciting pathway toward integrated multifunctional materials.

\section{Research Opportunities and Application Prospects}

The structural architecture of layered ferroelectrics unlocks unique research directions and application prospects. Their polarization is unusually robust to free charge carriers, allowing the coexistence of polar displacements with metallic conductivity and the formation of layer-confined two-dimensional electron gases (Sec.~\ref{sec:polar_metals}). The same structural motifs facilitate magnetic substitution without suppressing ferroelectricity, while the octahedral tilts and rotations often associated with polarization enable symmetry-allowed coupling to magnetic order via the Dzyaloshinskii–Moriya interaction. The octahedral tilts and rotations are also expected to support altermagnetic order with polarization-tunable spin splitting (Sec.\ \ref{sec:ME}). Finally, the quasi-two-dimensional crystal framework permits post-growth soft-chemistry manipulation, such as exfoliation, ion exchange, and topotactic redox transformations, offering routes to engineer new phases and functionalities beyond as-grown films (Sec.~\ref{sec:exfoliation}). These features make layered ferroelectrics a practical platform to co-engineer polarization, spin order, and itinerant carriers within one material scaffold.

\subsection{Resistance to charge doping: polar metallicity and 2D electron gases}\label{sec:polar_metals}

Ferroelectricity and metallicity are traditionally considered incompatible, as free carriers at the Fermi level screen long-range dipole interactions and hinder the rehybridization processes that enable polar distortions\cite{bhowalPolarMetalsPrinciples2023}. Moreover, ferroelectricity requires reversible switching of polarization under applied electric fields, which was deemed unfeasible in conducting systems. Nonetheless, recent work has established the existence of polar metals, materials that combine polar displacements with metallic conductivity\cite{Puggioni2018a, hickox-young2023, zhou2020b}. Theoretical studies suggest that polar displacements can persist if conduction electrons couple weakly to soft phonon modes\cite{puggioni2014a}, or can even be enhanced via screening charge rearrangement\cite{zhao2018a}. Polar metallicity was experimentally demonstrated in LiOsO\textsubscript{3}\cite{shi2013, xiaoElectricalDetectionFerroelectriclike2020b}, while switchable ferroelectricity was achieved in a metallic 2D electron gas hosted at Ca-doped SrTiO\textsubscript{3}\cite{brehin}. Therefore, it appears that combining ferroelectricity and metallicity is not only possible but also could be useful, as non-centrosymmetric metals exhibit a bulk Rashba effect, allowing for the creation of spin-polarized currents that potentially could be modulated electrically. Below, we examine these concepts in layered perovskite-based ferroelectrics, where polarization persists under substantial doping and coexists with layer-confined conduction, as shown by theory and experiment.

\begin{figure}[htb!]
  \centering 
  \begin{adjustbox}{width=0.88\textwidth, center}
    \includegraphics[width=0.88\textwidth,clip, trim=5 5 5 5]{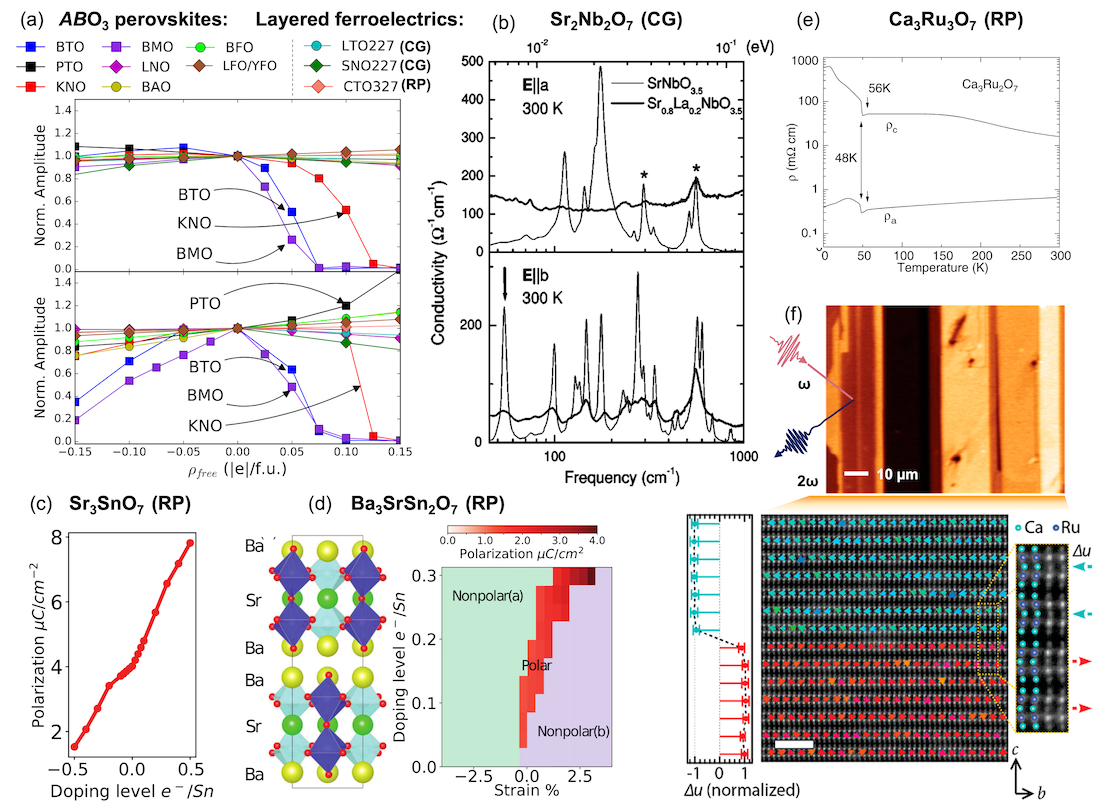}
  \end{adjustbox}
  \caption{\textbf{Persistence of polarization in layered ferroelectrics upon free-carrier doping.} 
(a)~First-principles calculations show that in contrast to conventional perovskite ferroelectrics, layered ferroelectrics (e.g., La\textsubscript{2}Ti\textsubscript{2}O\textsubscript{7} (LTO227), Sr\textsubscript{2}Nb\textsubscript{2}O\textsubscript{7} (SNO227), Ca\textsubscript{3}Ti\textsubscript{2}O\textsubscript{7} (CTO327)) retain significant polar displacements even under substantial electron or hole doping\cite{zhao2018a}. Adapted with permission from~\citenum{zhao2018a}. Copyright 2018 American Physical Society.  
(b)~Optical conductivity of undoped and La-doped (equivalent to electron doping) Sr\textsubscript{2}Nb\textsubscript{2}O\textsubscript{7} reveals persistent phonon modes and the soft mode of the ferroelectric parent phase in the doped compound, suggesting coexistence of metallicity and polar order\cite{kuntscher2004b}. Reproduced with permission from~\citenum{kuntscher2004b}. Copyright 2004 American Physical Society.  
(c)~In Sr\textsubscript{3}Sn\textsubscript{2}O\textsubscript{7}, polar displacements are theoretically predicted to be enhanced rather than suppressed by electron doping\cite{li2021a}. Reproduced with permission from~\citenum{li2021a}. Copyright 2021 American Physical Society.  
(d)~A doping-strain phase diagram of Ba\textsubscript{3}SrSn\textsubscript{2}O\textsubscript{7} shows that electron doping can induce and stabilize polarization in otherwise nonpolar phases\cite{li2021a}. Reproduced with permission from~\citenum{li2021a}. Copyright 2021 American Physical Society.  
(e)~Temperature-dependent resistivity in Ca\textsubscript{3}Ru\textsubscript{2}O\textsubscript{7} demonstrates anisotropic metallic behavior\cite{yoshida2004}. Reproduced with permission from~\citenum{yoshida2004}. Copyright 2004 American Physical Society.  
(f) SHG imaging confirms the presence of polar domains, and atomic-resolution STEM shows a 180$^\circ$ domain wall between two in-plane polarization directions in Ca\textsubscript{3}Ru\textsubscript{2}O\textsubscript{7} with measurable off-centering of Ca ions\cite{lei2018}. Adapted with permission from~\citenum{lei2018}. Copyright 2018 American Chemical Society.}
  \label{fig:polar_metals}
\end{figure}

\textbf{Robustness of polar displacements in layered ferroelectrics upon doping with free charge carriers.} First-principles studies have investigated how polar displacements respond to carrier doping in a variety of ferroelectrics\cite{zhao2018a}. While conventional perovskite ferroelectrics such as BaTiO$_3$ exhibit a rapid attenuation of polar distortions with increasing electron or hole concentration, several layered ferroelectrics display remarkable resilience (Fig.~\ref{fig:polar_metals}a). In particular, La$_2$Ti$_2$O$_7$ and Sr$_2$Nb$_2$O$_7$, both members of the Carpy–Galy family, as well as Ca$_3$Ti$_2$O$_7$, a Ruddlesden–Popper phase, maintain robust polar displacements even under substantial doping levels. Polar metallicity has also been theoretically predicted in related, yet experimentally unrealized Carpy–Galy compounds such as Bi$_5$Ti$_5$O$_{17}$\cite{filippetti2016} and Bi$_5$Mn$_5$O$_{17}$\cite{urru2020}. These predictions are supported by experimental observations in bulk Sr$_2$Nb$_2$O$_7$\cite{kuntscher2004b}, where optical conductivity measurements revealed the persistence of phonon modes, characteristic of the insulating ferroelectric phase, even after electron doping via La substitution or oxygen vacancies (Fig.~\ref{fig:polar_metals}b). In particular, the ferroelectric soft mode remains detectable in the doped compound, strongly suggesting that polar order coexists with metallic conductivity in these materials. Going one step further, first-principles calculations suggest that carrier doping can not only preserve but even enhance polar distortions in certain Ruddlesden–Popper phases. For instance, in Sr$_3$Sn$_2$O$_7$, electron doping strengthens the polar displacements\cite{li2021a} (Fig.~\ref{fig:polar_metals}c), and in Ba$_3$SrSn$_2$O$_7$, doping can even stabilize a polar phase in an otherwise centrosymmetric structure\cite{li2021a} (Fig.~\ref{fig:polar_metals}d).

\textbf{Polar domains uncovered in a metal.} A different manifestation of polar metallicity is seen in Ca$_3$Ru$_2$O$_7$, a bilayer Ruddlesden–Popper compound long known for its anisotropic metallicity and magnetic ordering\cite{yoshida2004}. Instead of introducing metallicity into a ferroelectric, this case starts from a correlated metal (Fig.~\ref{fig:polar_metals}e) and investigates whether polar distortions can coexist with itinerant carriers. Direct evidence for polar order in this compound comes from SHG imaging, which reveals ferroelastic and polar domains at room temperature (Fig.~\ref{fig:polar_metals}f). Furthermore, atomic-resolution STEM measurements uncover in-plane cation off-centering consistent with hybrid improper ferroelectricity, and even resolve 180$^\circ$ domain walls separating oppositely polarized regions (Fig.~\ref{fig:polar_metals}f).

\textbf{Why do layered ferroelectrics tolerate induced metallicity?} Several structural and mechanistic factors help explain the robustness of polarization in layered systems, even when free carriers are introduced. First, these compounds often exhibit high Curie temperatures (Sec.~\ref{sec:tc}), which reflects the fact that their ferroelectricity is stabilized not by local $B$-O hybridization as in classical perovskites, but by structural distortions on a larger unit-cell level. As a result, the ferroelectricity in these materials is inherently more robust against screening from mobile carriers. Second, the underlying polarization mechanisms differ significantly between conventional and layered ferroelectrics. Ruddlesden–Popper phases exhibit hybrid improper ferroelectricity (Sec.~\ref{sec:RP}), which is driven by a coupling between nonpolar modes, while polarization appears only as a secondary effect. Since these primary modes are not directly sensitive to electronic band filling, doping has a limited impact on the polar state. In the case of Carpy–Galy compounds, the situation is more nuanced. These are proper ferroelectrics, but the polarization does not originate from $B$-site off-centering, but rather from octahedral rotations and $A$-site cation displacements (Sec.~\ref{sec:CG}). As this mechanism does not rely on orbital re-hybridization, it remains largely intact even when $d$-electrons are introduced through doping. 

\begin{figure}[htb!]
  \centering 
  \begin{adjustbox}{width=0.88\textwidth, center}
    \includegraphics[width=0.88\textwidth,clip, trim=5 5 5 5]{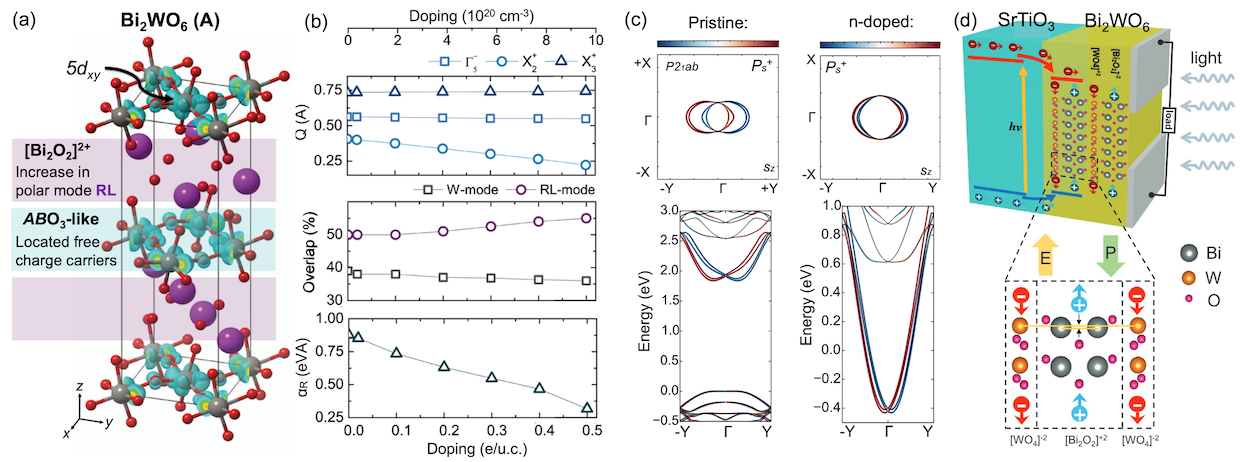}
  \end{adjustbox}
  \caption{
\textbf{Layer-selective carrier confinement in electron-doped Aurivillius ferroelectrics.}  
(a–c) Electron doping and Rashba splitting in Aurivillius Bi\textsubscript{2}WO\textsubscript{6}\cite{djani2019b}: (a) Charge-density of the conduction states in the electron-doped Bi$_2$WO$_6$, showing carrier confinement in the perovskite-like [WO$_4$]$^{2-}$ layers, while [Bi$_2$O$_2$]$^{2+}$ fluorite-like spacers remain mostly insulating. (b) Evolution of structural distortions as a function of doping reveals that while the total polar mode amplitude remains nearly unchanged, its perovskite-related component (W-mode) decreases and Bi\textsubscript{2}O\textsubscript{2}-related component (RL-mode) increases, preserving the overall amplitude of the polar mode. The Rashba coefficient decreases with doping but remains finite. (c) Spin-resolved band structures show retained Rashba splitting in the doped state, opening prospects for ferroelectric Rashba devices. Adapted with permission from~\citenum{djani2019b}. Copyright 2019 Springer Nature.  
(d) Schematic of photoexcited carrier separation mechanism in Bi\textsubscript{2}O\textsubscript{2}\cite{yang2020b}: photoexcited electrons from SrTiO\textsubscript{3} are injected into the perovskite-like layers of the Aurivillius film, enabling confined transport within a 2D potential defined by the layered structure. The spatial separation of electrons and holes across different layers reduces recombination and results in a giant enhancement in photocurrent compared to perovskite ferroelectrics. Reproduced with permission from~\citenum{yang2020b}. Copyright 2020 Wiley‐VCH GmbH.}
  \label{fig:BWO}
\end{figure}

\textbf{Creation of 2D electron gas in electron-doped Aurivillius compounds.} Compared to other layered ferroelectrics, Aurivillius phases have not traditionally been explored in the context of polar metallicity. This is likely due to the fact that their polarization mechanism, characterized by $B$-cation off-centering within the perovskite blocks closely resembles that of conventional ferroelectrics. However, recent studies suggest that charge doping in Aurivillius compounds can lead to unusual charge confinement (Fig.~\ref{fig:BWO}). 

Djani \textit{et al.}\ demonstrated through first-principles calculations that in Bi$_2$WO$_6$, conduction electrons introduced via doping predominantly localize within the perovskite-like blocks, while the fluorite-like [Bi$_2$O$_2$]$^{2+}$ layers remain largely insulating (Fig.~\ref{fig:BWO}a)\cite{djani2019b}. This spatial separation leads to the formation of a two-dimensional electron gas (2DEG) within the perovskite layers of the structure, similarly to confined conductivity reported in layered antipolar tungsten bronzes\cite{Nimoh2024}. Notably, despite the added carriers, the total polar distortion in Bi$_2$WO$_6$ remains relatively robust (Fig.~\ref{fig:BWO}b). By decomposing the polar mode, it was shown that while the displacements in the perovskites diminish (W-mode), the [Bi$_2$O$_2$]$^{2+}$ layers become more polar (RL-mode). Such coexistence of confined conduction and polarization opens up prospects for spintronic applications based on Rashba-type ferroelectrics, in which spin splitting of the conduction bands can be tuned via an external electric field. Although the splitting parameter decreases with increasing electron concentration (Fig.~\ref{fig:BWO}b), it remains finite (Fig.~\ref{fig:BWO}c). These results point to Aurivillius compounds such as Bi$_2$WO$_6$ as candidates for voltage-tunable spintronic devices.

\textbf{Layer-confined carriers enabling enhanced photoresponse.}  
The concept of spatially separated carrier confinement in layered ferroelectrics has also been experimentally harnessed to improve photoconductive performance. In Bi$_2$WO$_6$/SrTiO$_3$ heterostructures, Yang \textit{et al.}\cite{yang2020b} demonstrated that photoexcited electrons generated in the SrTiO$_3$ substrate are efficiently injected into the Bi$_2$WO$_6$ conduction bands of the Aurivillius film due to favorable band alignment. Crucially, the layered structure leads to natural confinement: electrons are localized within the perovskite-like blocks, while holes remain in the fluorite-like [Bi$_2$O$_2$] layers (Fig.~\ref{fig:BWO}d). This spatial separation suppresses electron–hole recombination and extends carrier lifetimes, resulting in an enhancement of the photocurrent. Compared to conventional perovskite ferroelectric BiFeO$_3$, the layered Bi$_2$WO$_6$ heterostructure exhibits orders-of-magnitude increase in photoresponse\cite{yang2020b}. The electronic confinement effects in Bi$_2$WO$_6$ were also experimentally investigated for negative differential resistance\cite{song2022}

\subsection{Coexistence with magnetic order: multiferroicity and (alter-)magnetoelectric coupling}\label{sec:ME}

The structural and electronic robustness of layered ferroelectrics not only enables coexistence with free charge carriers but also with magnetic ions, opening avenues toward multiferroicity. In many layered architectures, the presence of multiple perovskite layers between spacers allows selective magnetic substitution without disrupting the overall polar order. Moreover, the low symmetry and complex distortion patterns, often involving octahedral tilts and rotations, can give rise to antisymmetric exchange interactions such as the Dzyaloshinskii–Moriya interaction (DMI). These features facilitate direct coupling between polarization and magnetic order, making layered ferroelectrics exciting candidates for engineering multiferroicity with magnetoelectric coupling\cite{fiebig2016a,Spaldin2019}, or enabling polarization control of altermagnetic spin splitting\cite{smejkal2020,smejkal2022}. The sections below detail representative examples showing how polarization can control magnetic order and spin splitting in layered perovskite-based architectures.

\textbf{Multiferroic Aurivillius phases through targeted magnetic doping.}  
Just as Aurivillius phases served as a foundational platform for exploring ferroelectricity in layered oxides, they also became the subject of early experimental investigations into layered multiferroic thin films. Initial efforts focused on Fe-doped Aurivillius compounds within the Bi$_4$Bi$_n$Fe$_n$Ti$_3$O$_{3n+3}$ homologous family\cite{birenbaum2014a,faraz2015b,cao2017a,song2018a}, but the resulting magnetic phases were typically weakly magnetic or antiferromagnetic at low temperatures only. Subsequent doping attempts using cobalt\cite{keeney2012b} or terbium\cite{faraz2018a} showed inconsistent results, often confounded by secondary magnetic impurity phases, highlighting the challenge of controlling phase purity in partial ion substitution\cite{gradauskaite2017a,pradhan2019}.

\begin{figure}[htb!]
  \centering 
  \begin{adjustbox}{width=0.97\textwidth, center}
\includegraphics[width=0.97\textwidth,clip, trim=5 5 5 5]{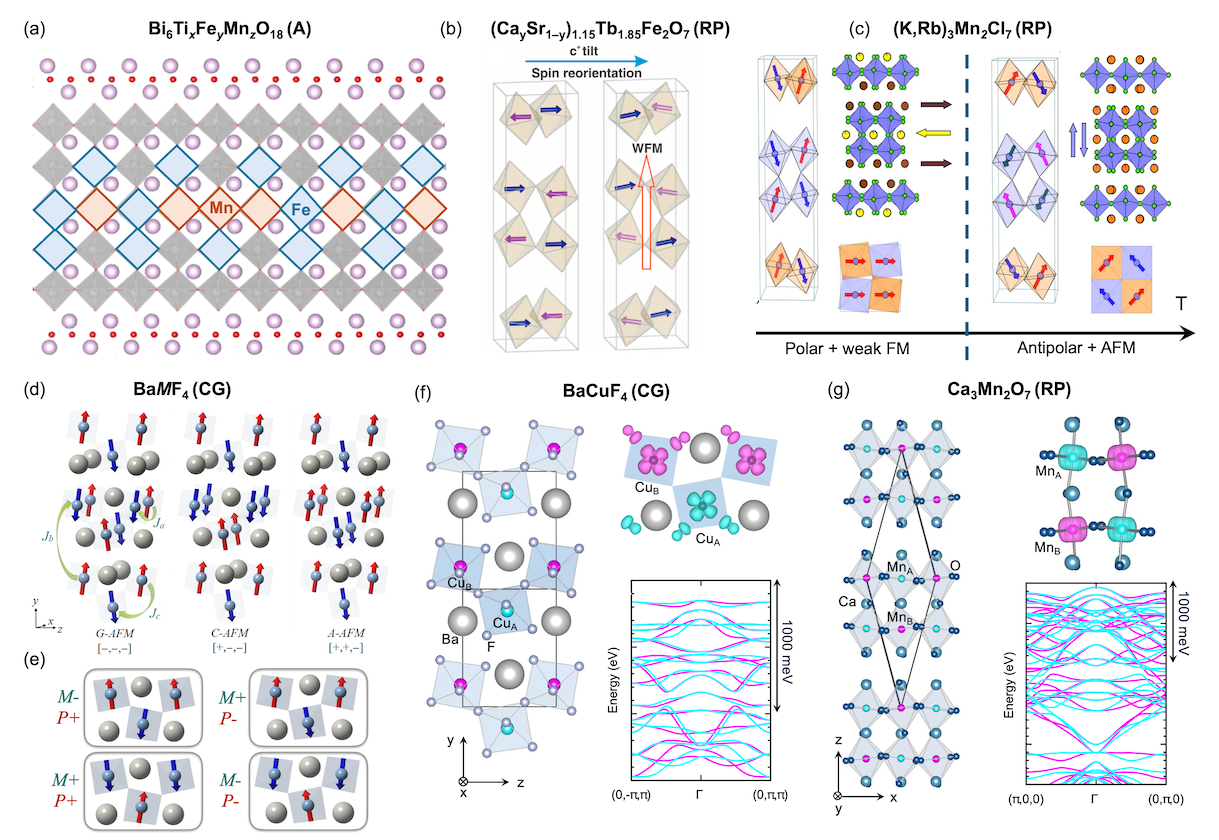}
  \end{adjustbox}
  \caption{\textbf{Multiferroic and magnetoelectric phenomena in layered ferroelectrics.}  
(a) The $n = 5$ compound Bi$_6$Ti$_x$Fe$_y$Mn$_z$O$_{18}$, exhibits room-temperature ferrimagnetism with magnetoelectric coupling\cite{keeney2013a,faraz2017a}. Magnetic order is stabilized despite the low doping level through preferential partitioning of Fe and Mn dopants into the central perovskite layers\cite{keeney2017a,moore2022b,colfer2024}.  
(b) Magnetoelectricity in (Ca,Sr)\textsubscript{1.15}Tb\textsubscript{1.85}Fe\textsubscript{2}O\textsubscript{7} is achieved by introducing a magnetic $B$-site sublattice and by using $A$-site doping to introduce tilts that give rise to spin canting and a weak ferromagnetic moment (wFM)\cite{pitcher2015a}. Reproduced with permission from~\citenum{pitcher2015a}. Copyright 2015 American Association for the Advancement of Science.  
(c) In quasi-2D halide (K,Rb)\textsubscript{3}Mn\textsubscript{2}Cl\textsubscript{7}, a thermal phase transition converts a polar–wFM state into an antipolar–antiferromagnetic state, enabling simultaneous thermal switching of magnetization and electric polarization\cite{zhu2024a}. Adapted with permission from~\citenum{zhu2024a}. Copyright 2024 Springer Nature.  
(d) Ba$M$F\textsubscript{4} compounds can adopt three different antiferromagnetic orders, among which either a weak antiferromagnetic moment (G-AFM)\cite{ederer2006} or a weak ferromagnetic moment (A-AFM)\cite{garcia-castro2018} can emerge via DMI, enabled by the presence of polarization. Adapted with permission from~\citenum{garcia-castro2018}. Copyright 2018 American Physical Society.  
(e) In BaCuF\textsubscript{4}, which is expected to stabilize in the A-AFM configuration, four multiferroic states arise from combinations of polarization and magnetization directions\cite{garcia-castro2018}. Reversing the polarization inverts the octahedral rotations, thereby switching the net magnetization. Adapted with permission from~\citenum{garcia-castro2018}. Copyright 2018 American Physical Society.  
(f-g) Altermagnetism in BaCuF\textsubscript{4} (f) and Ca\textsubscript{3}Mn\textsubscript{2}O\textsubscript{7} (g): nonrelativistic spin splitting associated with altermagnetism is revealed by the magnetization density around the magnetic ions and calculated electronic bandstructure\cite{smejkal2024}. Reproduced with permission from~\citenum{smejkal2024}.}
  \label{fig:ME}
\end{figure}

A key advancement came with the co-substitution of manganese and iron in Aurivillius phases. In particular, the $n = 5$ compound Bi$_6$Ti$_x$Fe$_y$Mn$_z$O$_{18}$ reached magnetic cation concentrations of up to 40\%, sufficient to stabilize room-temperature multiferroicity\cite{keeney2013a}, later proven to be a single-phase property\cite{schmidt2014a}. Magnetoelectric coupling was confirmed via lateral PFM, which revealed modified ferroelectric domain patterns upon application of a magnetic field\cite{keeney2013a,faraz2017a}. Atomic-resolution STEM studies\cite{keeney2017a} and first-principles calculations\cite{moore2022b} further showed that magnetic cations preferentially occupy the central perovskite layers, driven by elastic and electrostatic interactions. Additionally, increased octahedral tilting and reduced tetragonality in the central layers\cite{colfer2024} likely support the internal partitioning of magnetic dopants (Fig.~\ref{fig:ME}a), enhancing their connectivity and interactions despite relatively low overall magnetic ion concentrations.

\textbf{Hybrid improper ferroelectricity as a route to magnetoelectric multiferroics.}  
The discovery of hybrid improper ferroelectricity provided a mechanistic route to stronger magnetoelectric coupling in single-phase materials. A combination of nonpolar octahedral rotations and tilts can directly affect magnetic ordering via mechanisms such as DMI. And as more than one lattice distortion can switch the polarization in a hybrid improper ferroelectric, this also means that there are more routes to deterministically control magnetization. Following the initial theoretical proposals\cite{benedek2012a}, one of the first experimental demonstrations of such coupling in a hybrid improper system was reported by Pitcher \textit{et al.}\ in 2015\cite{pitcher2015a}. By carefully tuning the $A$-site composition to control the tilt geometry and introducing magnetic Fe on the $B$-site, they identified a crystal-chemically engineered multiferroic (Ca,Sr)\textsubscript{1.15}Tb\textsubscript{1.85}Fe\textsubscript{2}O\textsubscript{7} in the Ruddlesden--Popper phase that remains polar and weakly ferromagnetic above room temperature  (Fig.~\ref{fig:ME}b), as resolved by neutron powder diffraction. This study showcased how tilt-mode engineering, enabled by the geometric nature of polarization in hybrid improper ferroelectrics, can drive multiferroicity in layered oxides. A subsequent experimental work focused on Ca\textsubscript{3}Mn\textsubscript{1.9}Ti\textsubscript{0.1}O\textsubscript{7}, another hybrid improper ferroelectric where optical SHG signal associated with polarization showed a marked enhancement below the Néel temperature\cite{zemp2024}, indicating strong coupling between spin canting and ferroelectric polarization. 

Meanwhile, further exploration of magnetoelectric multiferroicity in hybrid improper ferroelectrics has been extended to other structural families. In Dion–Jacobson phases, a recent study demonstrated ion exchange of Li by Mn in polar Li$_2$SrTa$_2$O$_7$, resulting in MnSrTa$_2$O$_7$ with coexisting magnetic and ferroelectric order\cite{zhu2021}. Several theoretical predictions have also expanded the range of candidate materials, including anti-Ruddlesden–Popper structures, such as Eu$_4$Sb$_2$O \cite{markov2021}, and Ruddlesden–Popper chalcogenides and halides\cite{lu2023a}. More recently, in a quasi-2D halide system (K,Rb)\textsubscript{3}Mn\textsubscript{2}Cl\textsubscript{7}, a thermally driven phase transition between a polar weak ferromagnetic state and an antipolar antiferromagnetic state was demonstrated experimentally\cite{zhu2024a} (Fig.~\ref{fig:ME}c).

\textbf{Direct magnetoelectric coupling in Carpy--Galy fluorides.}  
The Ba$M$F$_4$ family ($M$ = Zn, Mg, Mn, Cu, Ni, Co, Fe) has been among the earliest studied multiferroic fluorides\cite{scott2011a}, with evidence of linear magnetoelectric coupling\cite{fox1980}. Structurally, these compounds belong to the Carpy–Galy family and were investigated soon after the discovery of magnetoelectricity in Cr$_2$O$_3$\cite{astrov}. They are collinear antiferromagnets and can stabilize in G-type, C-type, and A-type antiferromagnetic orders, with nearly degenerate energies (Fig.~\ref{fig:ME}d). Ferroelectricity additionally introduces spin canting along the $c$-axis via DMI, with distinct outcomes depending on the initial antiferromagnetic order. In G-AFM order, this spin canting induces a weak antiferromagnetic moment that couples directly to the polarization through octahedral rotations, as predicted for BaNiF$_4$\cite{ederer2006}. The same ordering is adopted in BaCoF$_4$, but experiments revealed that anisotropic strain can, in addition, stabilize a weak ferromagnetic (wFM) canting along the $x$-axis too\cite{borisov2016}. In contrast, A-AFM ordering leads directly to a wFM canting along $z$ in the presence of DMI, which was experimentally observed in BaCuF$_4$\cite{dance1981}. Recent first-principles studies\cite{garcia-castro2018} predict that in BaCuF$_4$, this wFM moment is directly coupled to the polarization via octahedral tilts characteristic of Carpy–Galy structures, which invert upon polarization reversal (Fig.~\ref{fig:ME}e). BaCuF$_4$ is predicted to exhibit a relatively high N\'eel temperature ($T_\mathrm{N} \approx 275$\,K), making it a rare fluoride multiferroic with near-room-temperature operation. Unfortunately, currently, there are no known naturally forming Carpy--Galy magnetic oxides that could be explored for such direct magnetoelectric coupling based on proper Carpy--Galy ferroelectricity.

\textbf{Altermagnetoelectric coupling in layered ferroelectrics.}  
Altermagnets are collinear antiferromagnets that break both time-reversal and certain crystallographic symmetries, leading to spin splitting in the electronic structure without a net magnetization\cite{smejkal2022}. In layered perovskite-based oxides, this unconventional order can be stabilized by electronic confinement that drives orbital ordering\cite{meier2025} or by structural confinement of octahedral rotation patterns\cite{smejkal2024}. In the layered ferroelectrics discussed in this review, the relationship between tilts and rotations associated with polarization and antiferromagnetism makes them excellent candidates to exhibit altermagnetism. In BaCuF$_4$ and Ca$_3$Mn$_2$O$_7$, both of which are layered ferroelectrics, first-principles calculations reveal $d$-wave altermagnetic spin order\cite{smejkal2024} (Fig.~\ref{fig:ME}f–g). Importantly, octahedral rotations in these materials can be directly controlled by the ferroelectric polarization switching, enabling an “altermagnetoelectric effect” in which reversal of the polarization switches the spin-splitting pattern. This effect provides a mechanism for voltage control of spin currents in antiferromagnets without relying on weak ferromagnetism or relativistic interactions, and opens new directions for spintronic applications in multiferroic materials with layered perovskite-based architectures.

\subsection{Two-dimensional nature: exfoliation and soft-chemistry-based phase transformations}\label{sec:exfoliation}

\begin{figure}[htb!]
  \centering 
  \begin{adjustbox}{width=0.88\textwidth, center}
\includegraphics[width=0.88\textwidth,clip, trim=5 5 5 5]{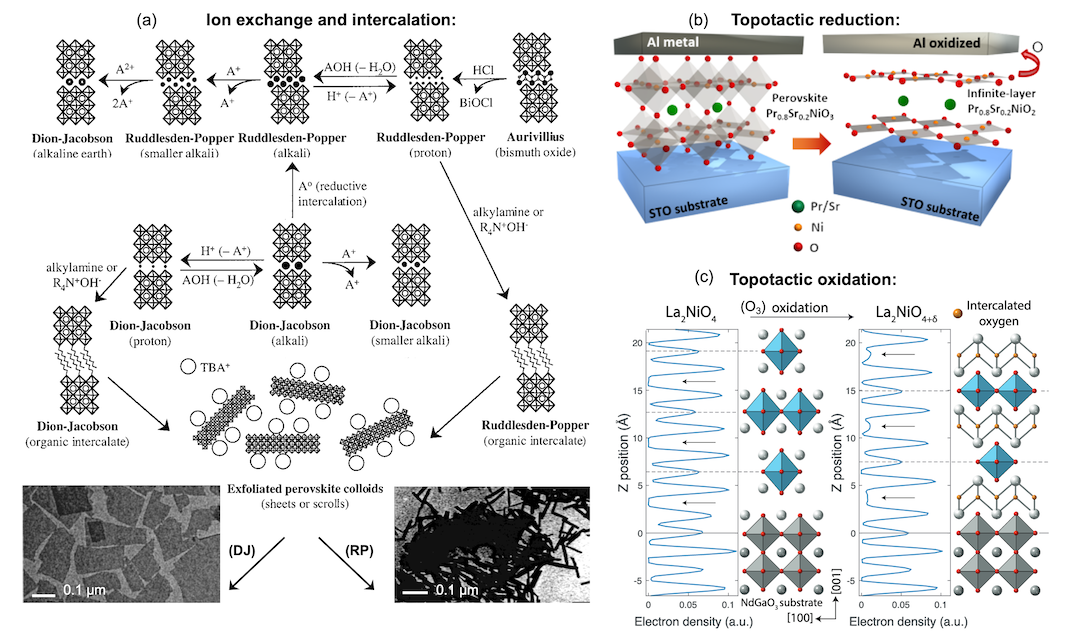}
  \end{adjustbox}
  \caption{\textbf{New directions for layered perovskite-based ferroelectrics through post-growth solid-state chemistry approaches.}   
(a) Layered perovskites can undergo multiple ion-exchange and intercalation reactions\cite{schaak2002a}, converting between structural families and enabling colloidal exfoliation of layered perovskite-based compounds into nanosheets\cite{schaak2000a} and scrolls\cite{schaak2000}. Adapted with permission from~\citenum{schaak2002a,schaak2000,schaak2000a}. 
Copyright 2000 and 2002 American Chemical Society.
(b) Topotactic reduction with an Al overlayer enables transformation of perovskite nickelates into superconducting infinite-layer phases\cite{zhang2024b}. Reproduced from \citenum{zhang2024b}.  
(c) Topotactic oxidation of Ruddlesden--Popper nickelates introduces interstitial oxygen into rock-salt layers, expanding the lattice and creating a new subclass of oxidized layered nickelates with features similar to the Aurivillius family\cite{segedin2025}. Reproduced from \citenum{segedin2025}.}
  \label{fig:exfoliation}
\end{figure}

While layered perovskite-based ferroelectrics are typically treated as three-dimensional solids, their intrinsic structural anisotropy gives them a pronounced two-dimensional character. In hybrid improper ferroelectric (Ca,Sr)$_3$Ti$_2$O$_7$ of the Ruddlesden–Popper family, through neutron scattering experiments, it has been shown that at high temperatures, the perovskite bilayers behave as isolated 2D slabs, with a 3D behavior emerging only at lower temperatures via dipole formation in the Ti–O planes\cite{kong2023a}. Similar behavior has been reported in Aurivillius phases, particularly those with even $n$, where weak interlayer coupling results in seemingly uncoupled buried in-plane-polarized domains across the film thickness\cite{campanini2019a}. These findings indicate that the ferroelectric behavior can reflect a fundamentally 2D-like character. Beyond electrostatics, the natural layering in these compounds allows for easier mechanical exfoliation along the spacer planes. Aurivillius thin films have been exfoliated using ultrasound to yield ferroelectric flakes of a sub-unit-cell thickness\cite{keeney2020c}. More recently, Dion–Jacobson CsPb$_2$Nb$_3$O$_{10}$ crystals were mechanically exfoliated using adhesive tape, producing few-monolayer flakes\cite{shimada2025}. While their interlayer spacing and exfoliation energies are higher than in typical van der Waals materials\cite{shimada2025}, the process remains accessible. These results highlight a path for producing 2D ferroelectric materials using simple mechanical techniques long established in the van der Waals 2D materials community.

\textbf{Soft-chemistry phase transformations, delamination, and nanosheet assembly.}  
Layered perovskite structures are also ideal candidates for soft-chemistry reactions, including ion-exchange, intercalation, and exfoliation through wet-chemical methods (Fig.~\ref{fig:exfoliation}a). We refer the readers to the comprehensive review  by Schaak \textit{et al.}\cite{schaak2002a} for more details. Importantly, Dion–Jacobson and Ruddlesden–Popper compounds can be also delaminated into nanosheets\cite{schaak2000a} or scrolls\cite{schaak2000} through chemical reactions. This is achieved through protonation of layered oxide phases, followed by treatment with Brønsted bases\cite{treacy1990}. The resulting nanosheets can be dispersed in colloidal suspensions and restacked via electrostatic self-assembly into multilayers\cite{schaak2000a}. The restacking process is self-limiting: absorption halts once surface charges are neutralized, allowing for well-defined thicknesses and unconventional stacking sequences inaccessible in bulk compounds. Chemical delamination of Aurivillius phases has also been demonstrated\cite{awaya2023}, although it resulted in Bi from the fluorite-like spacer layers becoming incorporated into the exfoliated perovskite nanosheets. Once separated into nanosheets, layered perovskites are attractive for applications such as photocatalysis, as they can be further functionalized through co-catalyst loading, doping, pillaring, etc.\cite{hu2020}.

\textbf{Blurring the boundary between solid-state chemistry approaches and physical vapor deposition of thin films.}  
Originally developed in wet-chemistry context, such phase transformations through solid-state chemistry reactions are now being adapted for \textit{in-situ} use during physical thin-film deposition. In particular, there has been a growing interest in the so-called topotactic phase transformations\cite{meng2023}, which are mediated by the ordered loss/gain and rearrangement of atoms, preserving thin-film epitaxial orientation. For example, topotactic reduction using a reactive aluminum capping layer allows conversion of perovskite nickelates into infinite-layer superconductors via oxygen removal and electron doping\cite{zhang2024b} (Fig.~\ref{fig:exfoliation}b). This is achieved simply by sputtering a few-nm-thick Al layer at elevated temperatures on the perovskite film\cite{zhang2024b} and gives identical results to those achieved through the chemical reduction with CaH\textsubscript{2}\cite{gutierrez-llorente2024}.

Conversely, topotactic post-growth oxidation with ozone has enabled insertion of interstitial oxygen into Ruddlesden–Popper nickelates, expanding their rock-salt layers and producing structures closer to Aurivillius-type phases\cite{segedin2025} (Fig.~\ref{fig:exfoliation}c). These post-growth modifications are transforming physical deposition processes into a dynamic platform for phase and property engineering. The emerging interplay between physical vapor deposition techniques and soft-chemistry approaches is expected to bring a shift in how layered perovskite-based ferroelectrics are designed, offering new possibilities for accessing metastable phases and reversibly tuning functionalities after growth.

\section{Conclusions and outlook}

Layered perovskite-based ferroelectrics, once considered structurally cumbersome and technologically outdated, are now reconsidered and started being explored more widely for their unconventional properties in thin films. The natural superlattice structure of these materials was long viewed as a limitation, making them seem difficult to synthesize in high-quality thin-film form. However, recent work shows that their intrinsic structural anisotropy can, in fact, aid their growth and provides better orientation control compared to conventional isotropic perovskites. Likewise, their in-plane polarization was once considered incompatible with standard ferroelectric capacitor architectures. Yet today, with new modes of electrical and optical readout, this characteristic in some cases has become an asset (e.g.\ absence of critical thickness for ferroelectricity) rather than a hindrance. 

This review aimed to present a unified picture of the layered perovskite-based ferroelectrics, including Aurivillius, Carpy–Galy, Ruddlesden–Popper, and Dion–Jacobson families. By covering developments in thin-film synthesis, characterization, and newly uncovered functionalities in thin films of these systems, this review highlights how the structurally distinct families of layered ferroelectrics share a surprising number of universal features. These include in-plane polarization with no critical thickness, high Curie temperatures, and common types of structural defects influencing ferroelectric domain formation. Layered compounds exhibit robust ferroelectricity, epitaxial compatibility with perovskite heterostructures, and non-trivial polar textures. Finally, their intrinsic layering makes them amenable to exfoliation and post-growth structural modifications through soft-chemistry approaches.

Recent discoveries in polar metallicity, multiferroicity, and altermagnetic ferroelectrics illustrate how the structural flexibility of layered ferroelectrics could enable novel and exciting effects at the frontier of condensed matter physics. Their compatibility with spin-based, optical, and electronic functionalities opens new directions that go well beyond classical applications like ferroelectric memories. In this spirit, we hope the review encourages a broader appreciation of layered ferroelectrics, not as isolated material systems, but as a versatile design framework, allowing us to go beyond the properties known for conventional perovskite ferroelectrics.

\bibliographystyle{apsrev4-2}
\bibliography{review}  

\end{document}